\documentclass[aip,jcp,reprint,floatfix]{revtex4-1}
\usepackage{amsmath}
\usepackage{graphicx}
\begin{document}

\title{
  Reversibility, Pattern Formation and Edge Transport in Active Chiral and Passive
  Disk Mixtures 
} 
\author{
  C. Reichhardt}
\affiliation{
Theoretical Division and Center for Nonlinear Studies,
Los Alamos National Laboratory, Los Alamos, New Mexico 87545, USA} 

\author{C. J. O. Reichhardt }
\altaffiliation{Corresponding author email: cjrx@lanl.gov}
\affiliation{
Theoretical Division and Center for Nonlinear Studies,
Los Alamos National Laboratory, Los Alamos, New Mexico 87545, USA}

\begin{abstract}
We numerically examine mixtures of circularly moving and passive disks as a function of density and active orbit radius.  For low or intermediate densities and/or small orbit radii, the system can organize into a reversible partially phase separated labyrinth state in which there are no collisions between disks, with the degree of phase separation increasing as the orbit radius increases.  As a function of orbit radius, we find a divergence in the number of cycles required to reach a collision-free steady state at a critical radius, while above this radius the system remains in a fluctuating liquid state.  For high densities, the system can organize into a fully phase separated state that is mostly reversible, but collisions at the boundaries between the phases lead to a net transport of disks along the boundary edges in a direction determined by the chirality of the active disk orbits.   We map the dynamic phases as a function of density and orbit radii, and discuss the results in terms of the reversible-irreversible transition found in other periodically driven non-thermal systems.  We also consider mixtures of circularly driven disks and ac driven disks where the ac drive is either in or out of phase with the circular motion, and find a rich variety of pattern forming and reentrant disordered phases.  
\end{abstract}

\maketitle

\section{Introduction}

There is a wide class of nonequilibrium systems that
undergo transitions from a non-fluctuating to a fluctuating state
as a function of increased
driving.
Examples include plastic depinning,
in which the transition is between
a non-fluctuating pinned or well-defined flowing channel state 
and a
strongly fluctuating 
plastically moving state \cite{1},
periodically sheared colloidal systems \cite{2},
active matter or self-driven systems \cite{3,4,5}, 
the transition to turbulence in driven liquid crystals
\cite{6}  
and the general class of systems that exhibit absorbing phase transitions \cite{7}. 
A particularly clear example of this
type of behavior appears in
periodically sheared dilute non-thermal colloidal particles,
where
at low shear or low colloid density,
the system organizes into
a reversible state
in which the particles return to the same positions
after each cycle,
but when the shear or density is above a threshold value,
the particles do not return and
undergo long time diffusion
\cite{2}. 
Corte {\it et al.} \cite{8}
showed that although
this system is always
initially in an irreversible or fluctuating state,
over many cycles
it organizes into either 
a reversible state or a steady irreversible state, with
the number of cycles or
amount of time
required 
to reach either of these phases diverging as a power law at a
critical drive.
In this case, reversible states correspond to the loss of
all
collisions between the particles.
Further studies 
of periodically driven systems have
revealed that
reversible to irreversible transitions  
can also occur in more strongly interacting systems
where the particles are always in 
contact, such as vortices in type-II superconductors \cite{9,10,11},
dense granular matter \cite{12,13},  and
periodically sheared amorphous solids \cite{14,15,16,17,18}.
In dilute colloidal systems, it was shown
that the particle configurations
at the reversible-irreversible transition
can exhibit hyperuniformity \cite{19,20,21}, and that
numerous
types of memory effects can arise \cite{22,23}.  

Active matter or self-propelled particles is
another example
of a nonequilibrium system
that often shows transitions from strongly
fluctuating disordered states
to weakly fluctuating or even nonfluctuating clustered states as a function 
of activity or density \cite{4,5,24,25,26}.
Studies of active matter frequently employ
particles 
that undergo run-and-tumble dynamics or driven diffusion \cite{26}.
Chiral active matter consisting of
circularly moving particles or spinners \cite{26,27}
is another class of active systems that can describe
biological circle swimmers \cite{26,27,28}, self-propelled colloids \cite{27,29,30}, 
active gears \cite{31,32,33}, interacting rotators \cite{34,35}, circularly driven particles \cite{36,37,38,39,40}, 
and collections of rotating robots \cite{41}.  

In this work we examine a binary assembly of non-polar disks
in which
half the disks undergo active
clockwise circular motion 
and the other half are passive. The system
is characterized by
the disk area coverage $\phi$
and the orbit radius $R$ of the active
disks.
We consider the limit of
no thermal
fluctuations,
so the system can be
trapped in a truly
non-fluctuating collision-free reversible state, and
we initialize the disks
in a non-overlapping dilute lattice.
At low disk densities,
the system remains in a reversible state
when the active radius is small enough,
but 
when the active radius increases above a threshold value,
a series of collisions between the active and passive
disks occurs
and the system organizes into a
phase separated cluster or labyrinth state in which
all collisions cease. 
The phase separation becomes stronger with increasing orbit radius,
and the active
disks perform circular orbits in the open regions
between clusters of passive disks. 
At large enough orbit radius,
the system remains in a permanently fluctuating state with
continuous collisions.
We show that the number of cycles or time
required to reach a reversible state
diverges at a critical
orbit radius
in a manner similar to that
found for
systems that exhibit random organization, where the 
time exponent  
is consistent with directed percolation or conserved directed percolation.
At higher disk densities,
the disks can organize
into pattern forming or completely phase separated states in which
reversible collisions
occur along the boundaries between the phases.
As the orbit radius increases, we find states in which
the motion of the bulk regions is reversible but a net transport
of
disks occurs along the phase boundaries in a direction that
is controlled by the 
chirality of the active
disks.
For very large orbit radii,
the system again enters a fluctuating liquid state.
We also examine samples in which the second species is not passive
but moves under a one-dimensional (1D) ac drive, and find
several distinct pattern forming states depending on whether
the 1D ac drive is in or out of phase with the circular motion of
the first
disk species.

\section{Simulation and System} 

We consider a two-dimensional system with periodic boundary conditions in the $x$ and $y$ directions containing
a total of $N$ disks that are each of radius $R_{d}$.
The system size is $L \times L$ with $L=36$, and
we take $R_d=0.5$.
The disk-disk interactions have a repulsive harmonic form, and the 
force between disks $i$ and $j$
is given by
${\bf F}^{ij}_{pp} = k(r_{ij} -2R_{d})\Theta(r_{ij} -2R_{d}){\bf \hat r}_{ij}$,
where $r_{ij} = |{\bf r}_{i} - {\bf r}_{j}|$,
$ {\bf \hat r}_{ij} = ({\bf r}_{i} - {\bf r}_{j})/r_{ij}$,
and $\Theta$ is the Heaviside step function.
As in previous works \cite{42,43},
we set the spring stiffness to $k = 50$, a value 
large enough to ensure
that there is
less than a one percent overlap between the
disks.
We characterize the system in terms of the
total 
area coverage  $\phi = N \pi R^{2}_{d}/L^2$, where 
a triangular solid forms when $\phi = 0.9$.
The dynamics of the disks 
are determined by the following overdamped equation of motion:
\begin{equation}
\eta \frac{d{\bf r}_i}{dt} = \sum_{j \neq i}^{N}{\bf F}_{pp}^{ij} + {\bf F}_{circ}^i 
\end{equation}
The first term is the
disk-disk interaction force and the second term
is the driving force ${\bf F}_{\rm circ}$
that
imposes a circular motion for $N_A$
of the disks and is
set to zero for the remaining $N_P=N-N_A$ disks.
Unless otherwise noted, we take $N_A=N_P=N/2$.
The circular ac drive has the form
${\bf F}_{\rm circ}={\bf F}^x_{c}+{\bf F}^y_c$, where
${\bf F}^{x}_{c} = A\sin(\omega t){\bf \hat x}$
and ${\bf F}^{y}_{c} = A\cos(\omega t){\bf \hat y}$.
We fix $\omega = 1\times 10^{-5}$, and vary the drive amplitude $A$. 
The damping constant $\eta$ is set equal to unity.
A single disk subjected only to the
circular drive
performs a clockwise circular
orbit with a radius $r_{a} = 2A$.     
The disks are initially placed in non-overlapping positions in a
diluted triangular lattice.
Typically, the system 
starts in a transient fluctuating state
and settles into a non-fluctuating or steady fluctuating state.  

\section{Results and Discussion}

\begin{figure}
\includegraphics[width=0.48\textwidth]{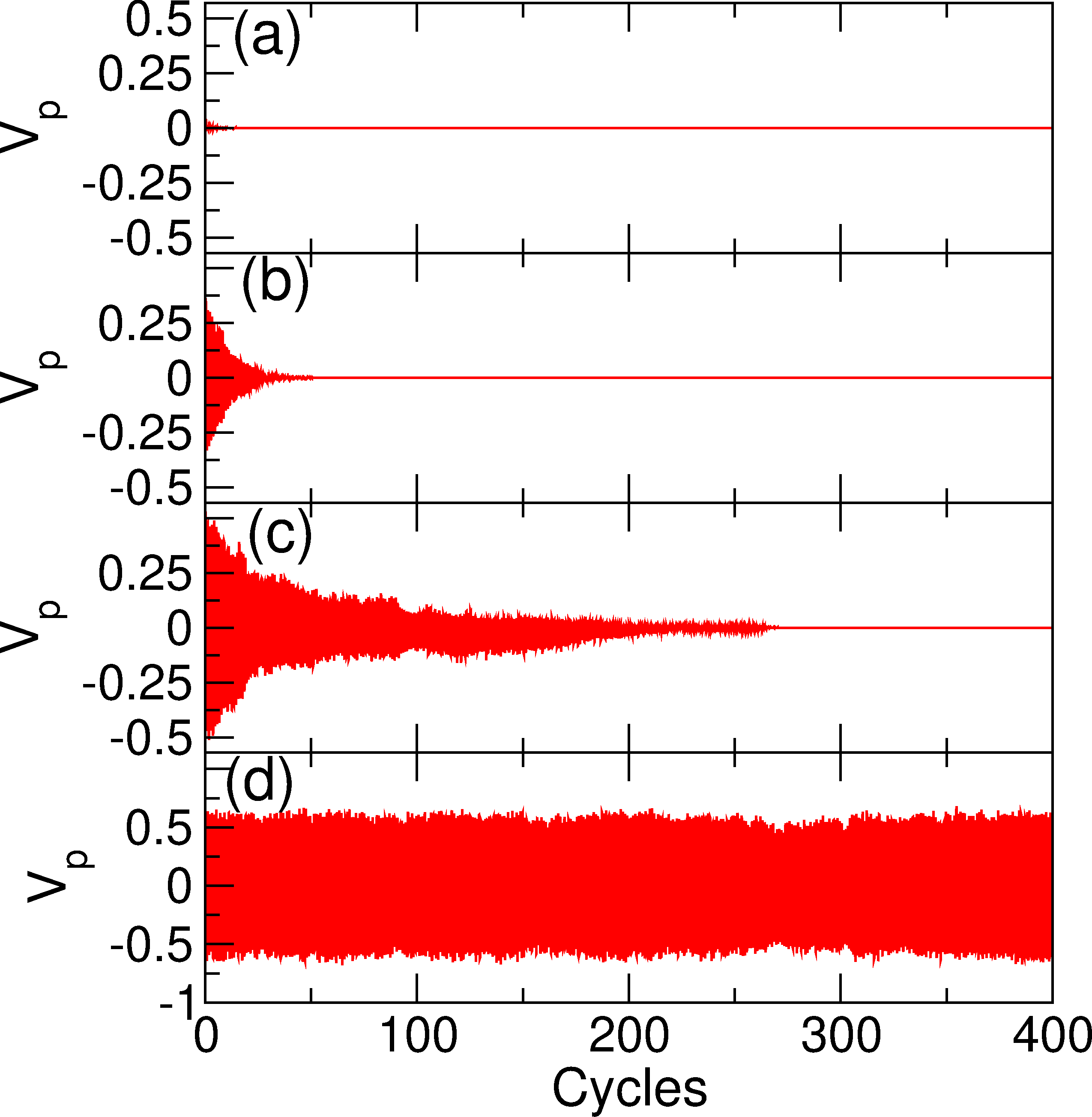}
\caption{
  $V_p$, the instantaneous $x$-component velocity of the passive disks,
  vs time in circular drive cycles
  for a binary assembly of disks
  in a system with a total disk density of $\phi=0.424$ where
  half of the disks
  are driven in a circular motion and the other half are passive.
  (a) At a circular drive amplitude of $A = 0.2$,
  the system
  is always in a collision-free state
  as indicated by the fixed value of $V_{p} = 0.0$.
  (b) At $A = 1.5$,
  $V_{p}$ is initially finite due to the occurrence of disk-disk
  collisions,
  but drops to zero
  after 30 cycles.
  (c) At $A = 2.2$, the system requires 250 cycles to reach a
  collision-free state with $V_p=0$.
  (d) At $A = 2.75$,
  the system remains in a fluctuating state with continuous
  disk-disk collisions.  
}
\label{fig:1}
\end{figure}

We first consider a system at a density of
$\phi = 0.424$ containing active disks labeled species 1 and
passive disks with $A=0$ labeled species 2.
We measure the $x$-component of the
instantaneous velocity of the passive species
$V_{p}=N_p^{-1}\sum_{i}^{N}\delta(A)({\bf v}_i \cdot {\bf \hat x})$.  
If the active
disks collide with the passive disks,
$V_p$ will have a finite instantaneous value
that fluctuates around zero;
however, if there are no collisions,
$V_{p} = 0.0$.
If the
system organizes over time
into a state where there are no collisions,
then $V_{p}$ will initially fluctuate around zero but will converge
over a number of cycles to the fixed value
$V_{p} = 0.0$.
In Fig.~\ref{fig:1}(a) we plot $V_{p}$ versus time in circular drive
cycles for the small circular drive amplitude of
$A = 0.2$.
Here we immediately obtain
$V_{p} = 0.0$
since the circular orbits are so small that
disk-disk collisions never occur.
At $A=1.5$
in Fig.~\ref{fig:1}(b),
$V_p$ initially has a finite value
with large oscillations, produced when a
large number of collisions
occur between the active and passive
disks,
but
$V_{p}$ gradually decreases and reaches zero after 30 cycles,
indicating that the system has organized
into a collision-free state.
In Fig.~\ref{fig:1}(c)
at 
$A = 2.2$,
it
takes 250 cycles to reach a collision-free state,
while at $A=2.75$ in
Fig.~\ref{fig:1}(d),
the system does not converge to a collision-free state but
remains in a permanently fluctuating state.

\begin{figure}
\includegraphics[width=\columnwidth]{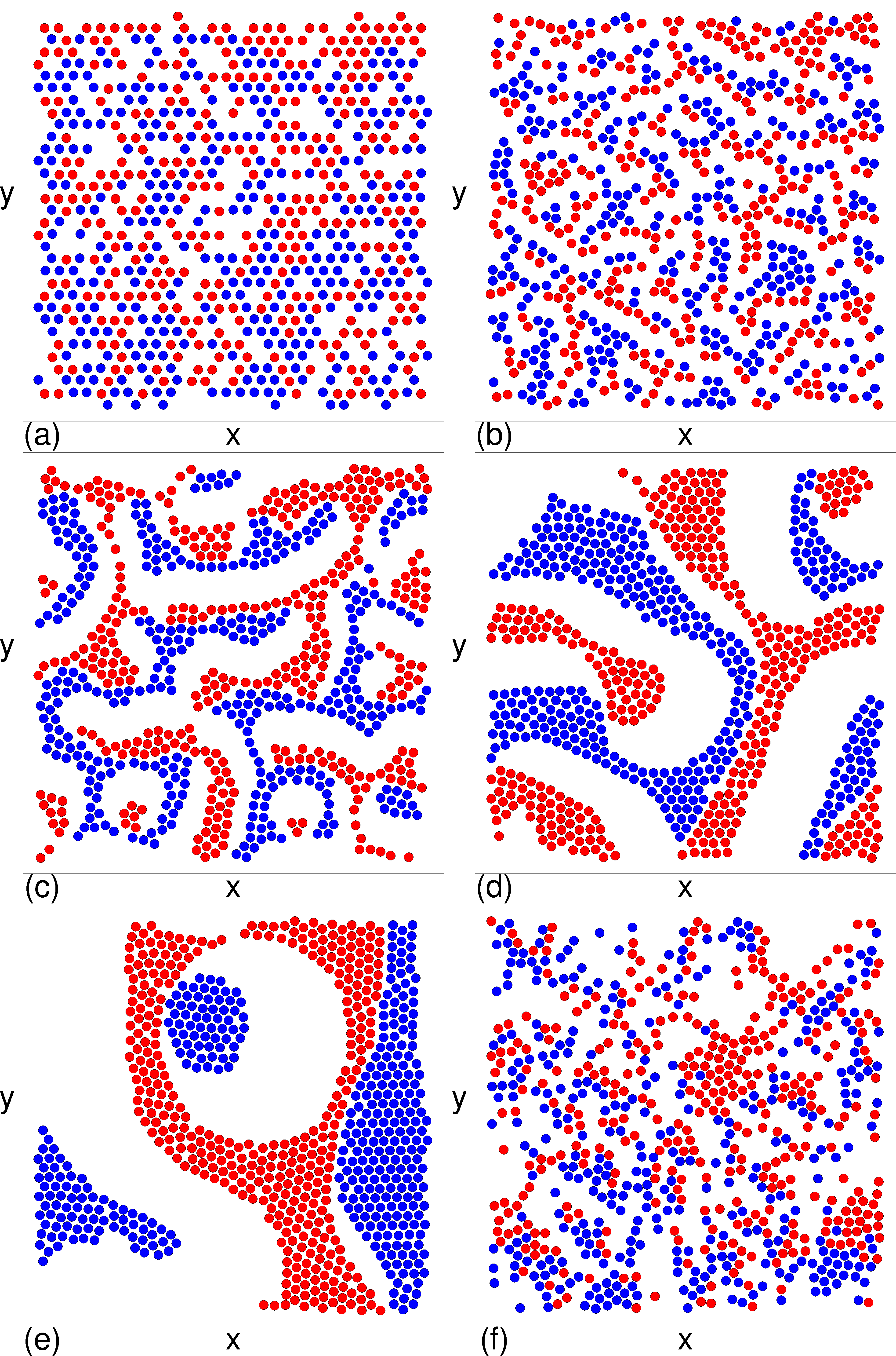}
\caption{Snapshot of instantaneous positions for the active (blue)
  and passive (red) disks for
  the system in Fig.~\ref{fig:1}
  at $\phi = 0.424$.
  (a) At $A = 0.1$,
  the system is always 
  in a collision-free state
  since the orbits of the active disks
  are too small for disk-disk collisions to occur,
  and there is no phase separation.
  (b) The collision-free state that forms for $A = 0.5$ after
  several cycles.
  (c) The collision-free state at $A = 0.75$ where
  we observe a partially phase separated labyrinth configuration.
  (d) The collision-free state at $A = 1.5$,
  where the widths of the labyrinths are larger. 
(e) The collision-free reversible state at $A = 2.3$ where the
  phase separation is the strongest.
  (d) At $A = 2.75$, the system remains in a permanently
  fluctuating state with collisions between the
passive and active disks.
}
\label{fig:2}
\end{figure}

In Fig.~\ref{fig:2}(a) we show the positions of the passive
and active
disks at $A = 0.1$ where the system
is always in a collision-free state since the
active disk orbits are too
small to produce disk-disk collisions.
Here the disks remain in their initial mixed state configuration. 
Figure~\ref{fig:2}(b) illustrates the reversible state
at $A = 0.5$ where the
disks are initially in
a fluctuating state and 
then settle into a collision-free state in which
there is some local clustering of both species.
At $A=0.75$ in the reversible state,
Fig.~\ref{fig:2}(c) shows
that the system forms a phase separated labyrinth pattern
containing
local dense regions
in which
disks of the same species 
are almost touching each other.
As $A$ increases,
the width of the labyrinths in the reversible state increases,
as shown in Fig.~\ref{fig:2}(d) at $A = 1.5$ 
for the system in Fig.~\ref{fig:1}(b) which requires
30 cycles to reach a reversible  state.
In Fig.~\ref{fig:2}(d) we illustrate
the reversible state at $A = 2.3$ which took 1050 cycles to
form.
This value of $A$ is just below
the critical value of $A_{c}$, above which
the system remains in a permanently fluctuating state.
We find the greatest amount of phase separation just below $A_c$ along with
considerable six-fold ordering within each phase, indicating that the
local density of the
dense regions is close to the solidification density of $0.9$.  
Figure~\ref{fig:2}(f)
shows a representative configuration in the fluctuating state above $A_c$ for
$A = 2.75$,
where the clustering is lost and the disk configurations are rapidly changing.

\begin{figure}
\includegraphics[width=\columnwidth]{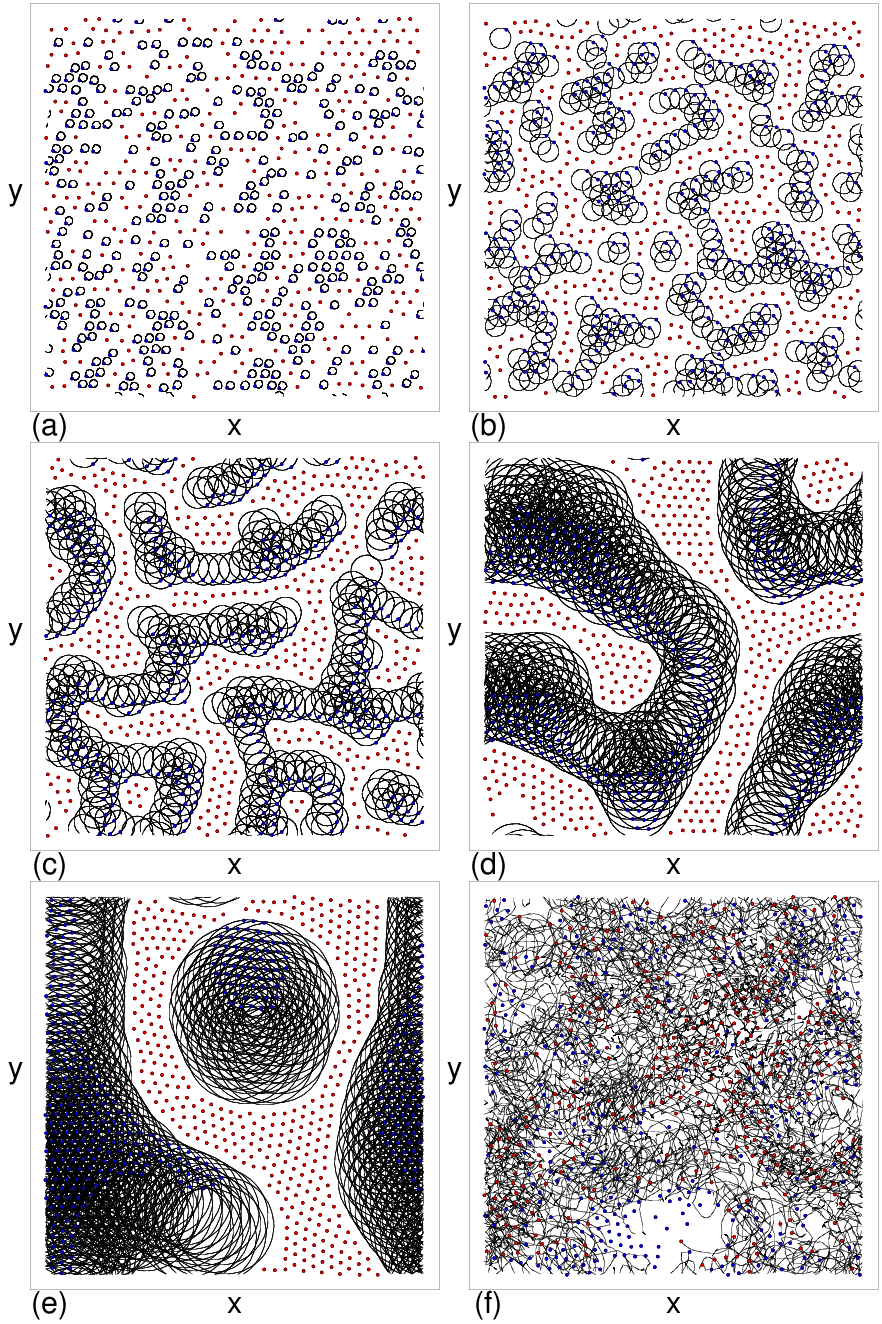}
  \caption{Snapshot of disk positions and trajectories
    in the collision-free state for
    the active (blue) and passive (red) disks
    in the system in Fig.~\ref{fig:2} at $\phi = 0.424$. 
    For clarity, the disks are drawn at one-third their actual size.
    (a) $A = 0.2$.
    (b) $A = 0.5$.
    (c) $A = 0.75$.
    (d) $A = 1.5$.
    (e) $A = 2.3$.
    (f) The trajectories of only the passive disks
    in the fluctuating state at $A = 2.5$,
    where
    the passive disks are undergoing collisions with the active disks.  
}
\label{fig:3}
\end{figure}

In Fig.~\ref{fig:3}(a-e) we highlight
the trajectories of the active disks during
a fixed number of cycles in the collision-free state
for the system in Fig.~\ref{fig:2} at $\phi_ = 0.424$.
At $A=0.2$, where Fig.~\ref{fig:1}(a) shows that the system immediately
enters a collision-free state, Fig.~\ref{fig:3}(a) indicates that the circular
orbits of the active disks do not overlap.
In Fig.~\ref{fig:3}(b) at
$A = 0.5$,
the motion
occurs in 
filaments with overlap in the orbits. In the reversible state 
the trajectories trace out the same path on every cycle.
At $A=0.75$ in Fig.~\ref{fig:3}(c),
the passive disks
are confined to regions
in which no active disk motion occurs.
This demonstrates
that the
empty spaces which appear in
Fig.~\ref{fig:2}(c) are periodically
occupied by the active disks during each drive cycle.
In Fig.~\ref{fig:3}(d),
the
$A = 1.5$ system from Fig.~\ref{fig:2}(d)
exhibits orbits that are much wider.
Figure~\ref{fig:3}(e) again shows
that in the $A=2.3$ system,
the empty regions found in
Fig.~\ref{fig:2}(e) correspond to locations through which the
active disks pass during each cycle.
The circular region of active disks surrounded
by passive disks in the upper portion of Fig.~\ref{fig:2}(e) is
actually rotating  
as a single group, as shown in Fig.~\ref{fig:3}(e).
There are no trajectory lines associated with the passive disks in the
reversible states since the passive disks are completely at rest.
This is in contrast to the behavior in the irreversible state, as
shown in 
Fig.~\ref{fig:3}(f) at $A=2.5$ where we highlight
the trajectories of only the passive
disks.
The system is in a permanently fluctuating state,
and the passive
disks are undergoing random motion that over
time produces a diffusive behavior.

\begin{figure}
\includegraphics[width=\columnwidth]{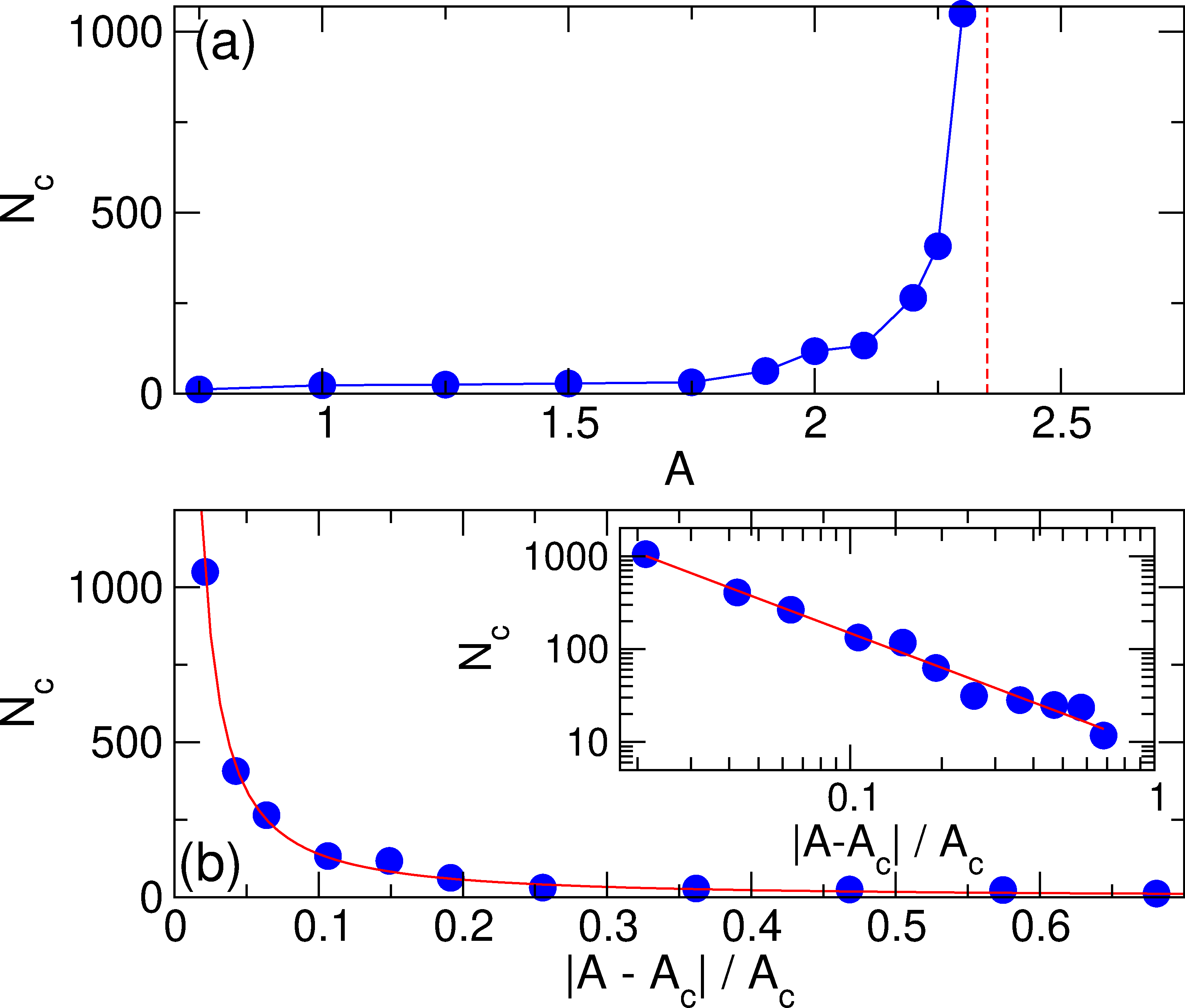}
  \caption{(a) The number of cycles $N_c$
    required to reach a reversible state vs $A$
    for the system in Figs.~\ref{fig:1} and \ref{fig:2} at 
    $\phi = 0.424$.
    $N_c$ diverges
    near $A_{c} = 2.35$, as indicated by the dashed line. 
    (b) $N_c$
    vs $(|A - A_{c}|/A_{c})$.  The solid line is a power law fit 
    to the form $N_c \propto (|A - A_{c}|/A_{c})^{-\nu}$
    with $\nu = 1.236 \pm 0.058$.
    The inset shows the same plot on a log-log scale. 
}
\label{fig:4}
\end{figure}

In Fig.~\ref{fig:4}(a) we plot the number of cycles $N_c$
required to reach the reversible state versus $A$ for the system 
in Figs.~\ref{fig:1} and \ref{fig:2} at $\phi = 0.424$.
There 
is a divergence in $N_c$ near $A_{c} = 2.35$,
as indicated by the vertical dashed line.
In Fig.~\ref{fig:4}(b) we show $N_c$
versus $|A - A_{c}|/A_{c}$, where the solid line is a power law fit
to the form $N_c \propto (|A - A_{c}|/A_{c})^{-\nu}$
with $\nu = 1.236 \pm 0.058$.
The inset shows the same data on a log-log scale. 
In two dimensional (2D) simulations of periodically sheared dilute colloids,
the number of shearing cycles needed to reach a reversible state 
also exhibits a power law divergence at a critical shear amplitude
with an exponent of $\nu = 1.33$,
while in three dimensional dilute colloid experiments,
the power law divergence has an exponent of
$\nu = 1.1$ \cite{8}.  
Other computational works for 2D
periodically driven dilute disk systems systems give $\nu = 1.26$ \cite{44}. 
Experiments in periodically driven vortex systems
produce the value $\nu = 1.3$ \cite{10}, while more recent simulations
of periodically driven skyrmions and vortices
show that $\nu \approx 1.3$ on the reversible side of the transition
\cite{45}.  
In contrast, 2D simulations of amorphous solids
that are in a strongly jammed state \cite{16} give
$\nu  = 2.53$ and $\nu = 2.4$,
suggesting that these amorphous jammed
systems fall into a different universality class.  
In principle, there can be another diverging time scale on
the irreversible side of the transition corresponding to the
time required for the system to reach
a steady, rather than a transient, fluctuating state;
however, the data on the irreversible
side of the transition in our system
is much harder to fit, so we concentrate 
on the behavior as the transition is approached
from the reversible side.
Our results are consistent with the
idea that the system undergoes an absorbing phase transition, 
since all collisions disappear on the reversible side of the
transition.
Absorbing phase transitions in 2D can fall into
the directed percolation universality class
with $\nu =1.30$ or the conserved directed percolation universality class
with $\nu = 1.23$ \cite{7}. 
Our results are not accurate enough to distinguish
between these two scenarios;
however, in our system the
number of disks is conserved,
and the exponent
we obtain is closer to that expected for conserved directed percolation.

\begin{figure}
\includegraphics[width=\columnwidth]{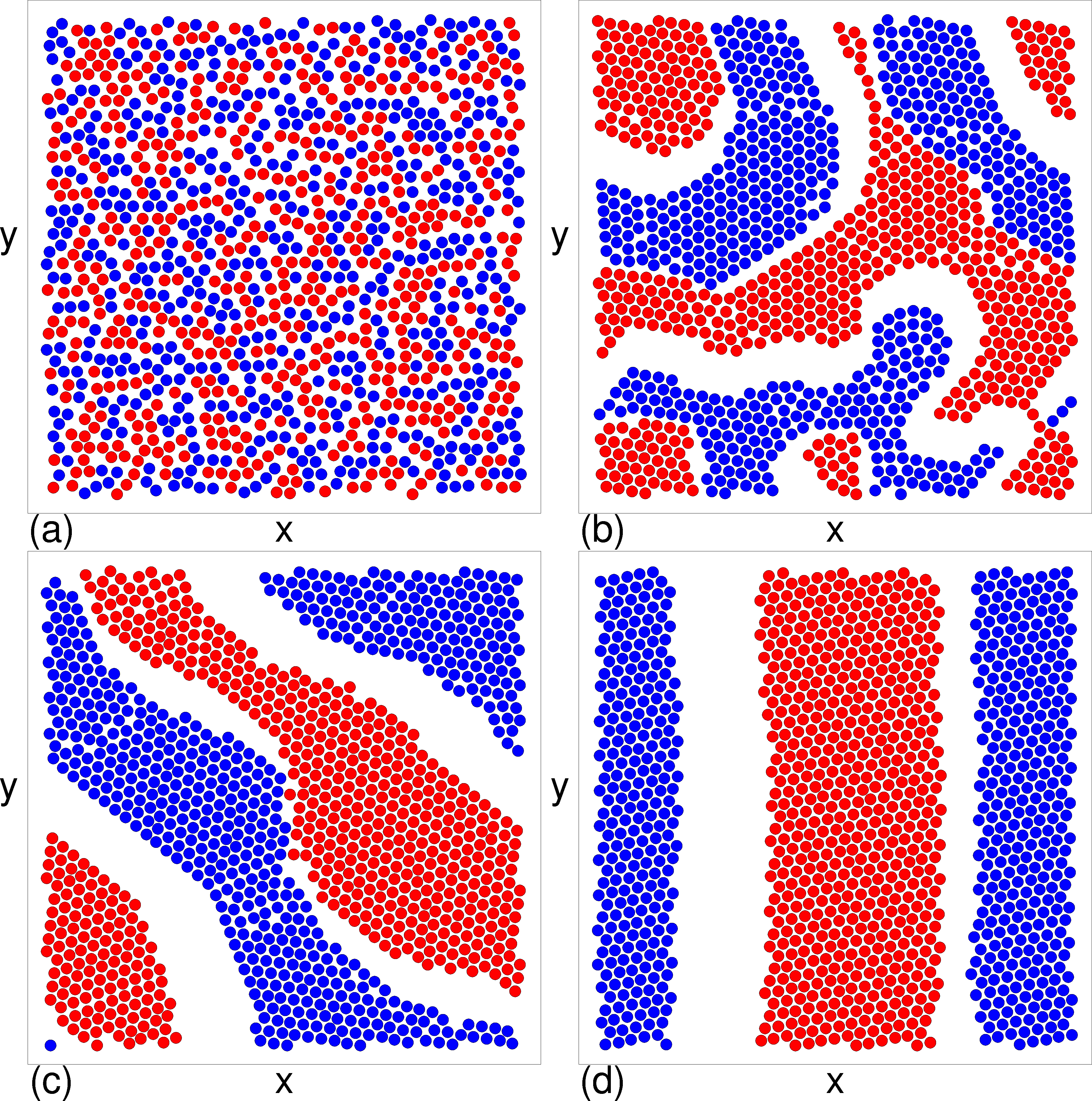}
\caption{
  Snapshot of instantaneous positions for the active (blue) and passive (red)
  disks
  at $\phi = 0.67$.
  (a) At $A = 0.1$ the system organizes to a collision-free state. 
(b) At $A = 0.75$ we find a reversible collision-free labyrinth state.  
  (c) At $A = 2.0$, the system is phase separated but there are
  reversible collisions along the phase boundaries.
  (d) At $A = 2.5$,
  the system is completely phase separated and there are reversible
collisions along the phase boundaries.  
}
\label{fig:5}
\end{figure} 

For $\phi < 0.5$ we find three well defined phases
as a function of varied $A$ and $\phi$.
At low $\phi$, as $A$ increases
the system either remains in an
initially reversible configuration, 
enters a reversible pattern forming state,
or exhibits a permanently fluctuating state.
For $\phi > 0.5$ we observe several new sub-phases,
including a pattern forming state and
a completely phase separated 
state in which {\it reversible} collisions
occur between some of the active and passive disks.
We also find a state in which
the system is reversible in the bulk but exhibits
a net transport of passive disks along the boundary of the
phase separated regions, meaning that
the
system is irreversible only along these boundary edges.  
In Fig.~\ref{fig:5}(a) we show the disk positions
at $\phi = 0.67$ and $A = 0.1$ where the 
disks have organized to a reversible
collision-free state, while in
Fig.~\ref{fig:5}(b) at the same disk density and
$A = 0.75$,
a fluctuation-free reversible labyrinth state
appears.
At $A=2.0$, illustrated in
Fig.~\ref{fig:5}(c),
even though the motion remains reversible,
collisions occur between the active
and passive disks along the boundaries of the phase separated regions.
In Fig.~\ref{fig:5}(d)
at $A = 2.5$,
the system is
completely phase separated,
collisions between disks at the phase boundaries continue to occur,
and the motion is still reversible.
At this density of $\phi=0.67$,
for $3.0 \leq A < 3.75$
the disks are
completely phase separated but
an irreversible flow of disks occurs along the phase boundaries that
generates a net edge current,
while for $A > 3.75$, the system
enters a disordered state
in which all the
disks
can diffuse.

\begin{figure}
\includegraphics[width=\columnwidth]{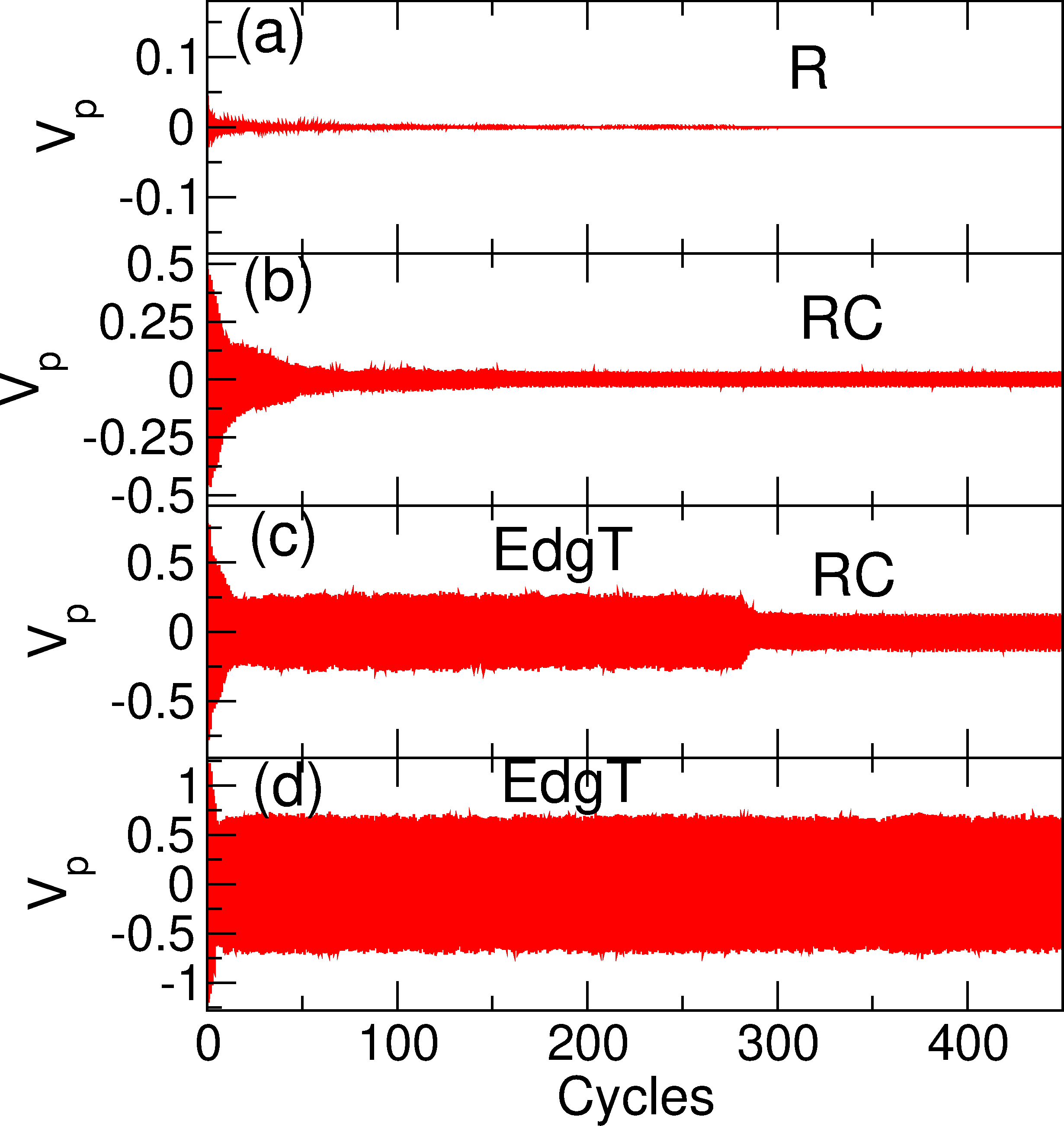}
\caption{
  $V_{p}$ vs time in circular drive cycles
  for the system in Fig.~\ref{fig:5} at $\phi = 0.67$.
  (a) $A = 0.1$,
  where
  the system forms a pattern forming reversible state with no collisions.
  (b) $A = 2.0$, where the system
  organizes into a completely phase separated
  state with reversible collisions (RC).
  (c) $A = 2.5$, where the system initially
  enters
  a completely phase separated state with edge transport (EdgT) but
  then transitions into a state with reversible collisions as
  illustrated in Fig.~\ref{fig:5}(d). 
(d) $A=4.0$, where edge transport occurs.
}
\label{fig:6}
\end{figure} 

In Fig.~\ref{fig:6} we
plot $V_{p}$ versus time in circular drive cycles
for the system in Fig.~\ref{fig:5} at $\phi = 0.67$.
At $A=0.1$, shown in
Fig.~\ref{fig:6}(a),
there is a short transient time
interval
during which disk
collisions and rearrangements occur before the system settles into a 
reversible collision-free state with $V_{p} = 0.0$,
as illustrated in Fig.~\ref{fig:5}(a). 
In Fig.~\ref{fig:6}(b)
at $A = 2.0$,
the system is in a disordered state
for the first 100 cycles,
and it then settles
into 
a completely phase separated reversible state
in which $V_{p}$
remains finite due to the
reversible collisions occurring
along the phase boundaries.  We call this state
the reversible collision (RC) phase.
At $A=2.5$, plotted in 
Fig.~\ref{fig:6}(c),
the system initially reaches
a phase separated state
in which irreversible
transport of disks occurs around the edges of the phase boundaries,
but after 275 cycles the system settles into a RC state as shown
in Fig.~\ref{fig:5}(d).
When the edge transport is occurring, $V_p$ exhibits large fluctuations 
but remains centered around zero
since the edge transport runs in
the positive $y$ direction on one side of the phase boundary and
in the negative $y$ direction on the other side. 
In Fig.~\ref{fig:6}(d)
at $A = 4.0$, the edge transport regime remains stable,
the system never becomes fully reversible, and the
fluctuations in $V_p$ are larger.
At higher drives (not shown),
the system enters
a liquid
state and the fluctuations in $V_{p}$
are even larger.

\begin{figure}
\includegraphics[width=\columnwidth]{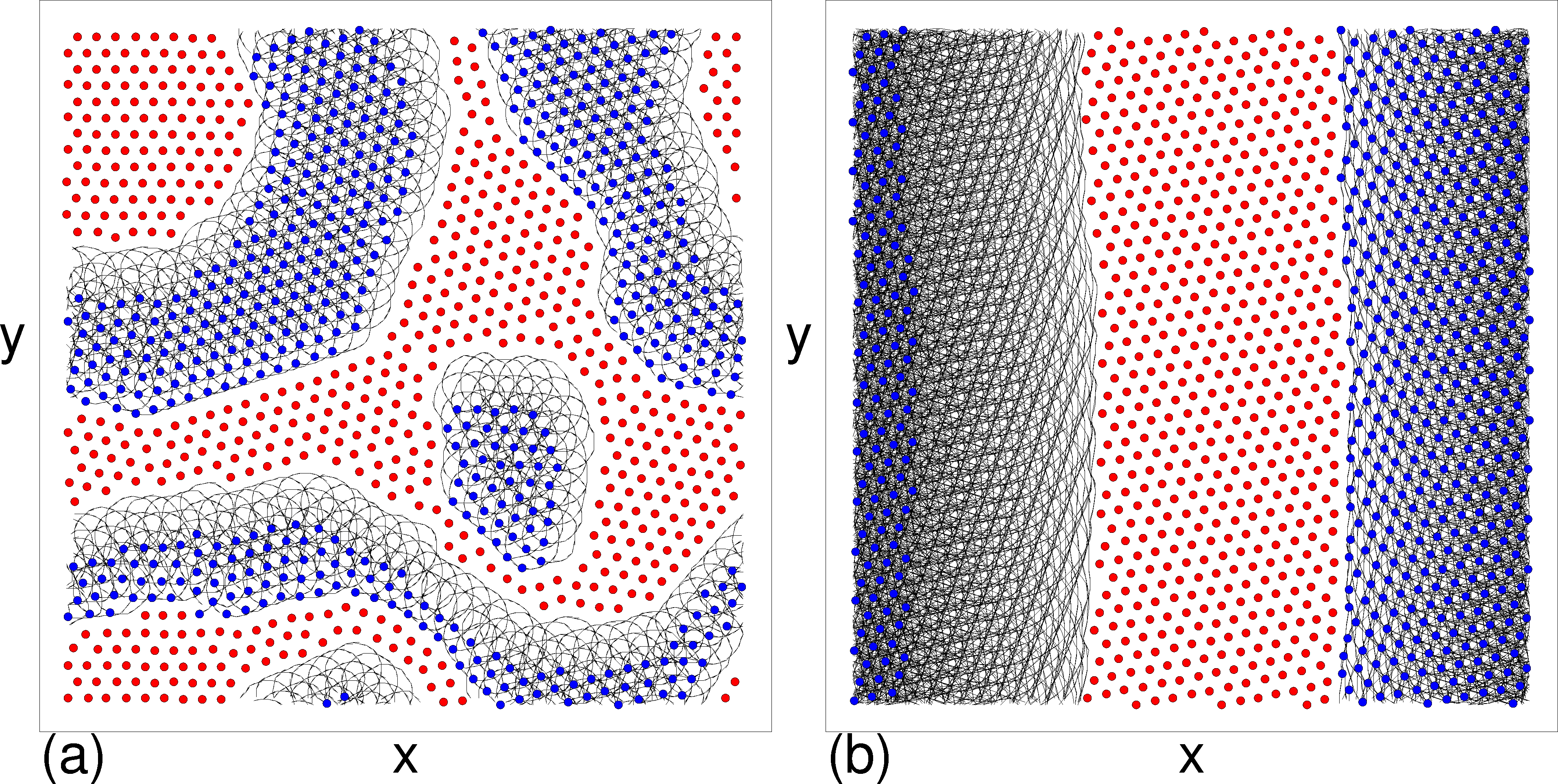}
\caption{Snapshot of disk positions and trajectories in the
  $\phi=0.67$ system for the active (blue) and passive (red) disks.  For clarity
  the disks are drawn at one-half their actual size.
  (a) $A = 1.5$, where there are no collisions.
  (b) $A = 2.5$, where reversible collisions occur.  
}
\label{fig:7}
\end{figure}

In Fig.~\ref{fig:7}(a) we show
the trajectories of the active
disks for
a reversible motion state at $A=1.5$
for the $\phi=0.67$ system from
Fig.~\ref{fig:6}.
In this reversible state, disks never come into contact with each other.
At $A=2.5$ in the same system,
Fig.~\ref{fig:7}(b) shows
that
reversible collisions now occur
between the active and passive disks
along the boundaries between the two species.
The large regions that are not occupied by passive disks are entirely
filled by the motion of the active disks.
Along the edges of the passive disk regions, passive disks
undergo small periodic orbits due to
collisions with the active disks, whereas
in the bulk of the passive disk regions,
little or no motion occurs.

\begin{figure}
\includegraphics[width=\columnwidth]{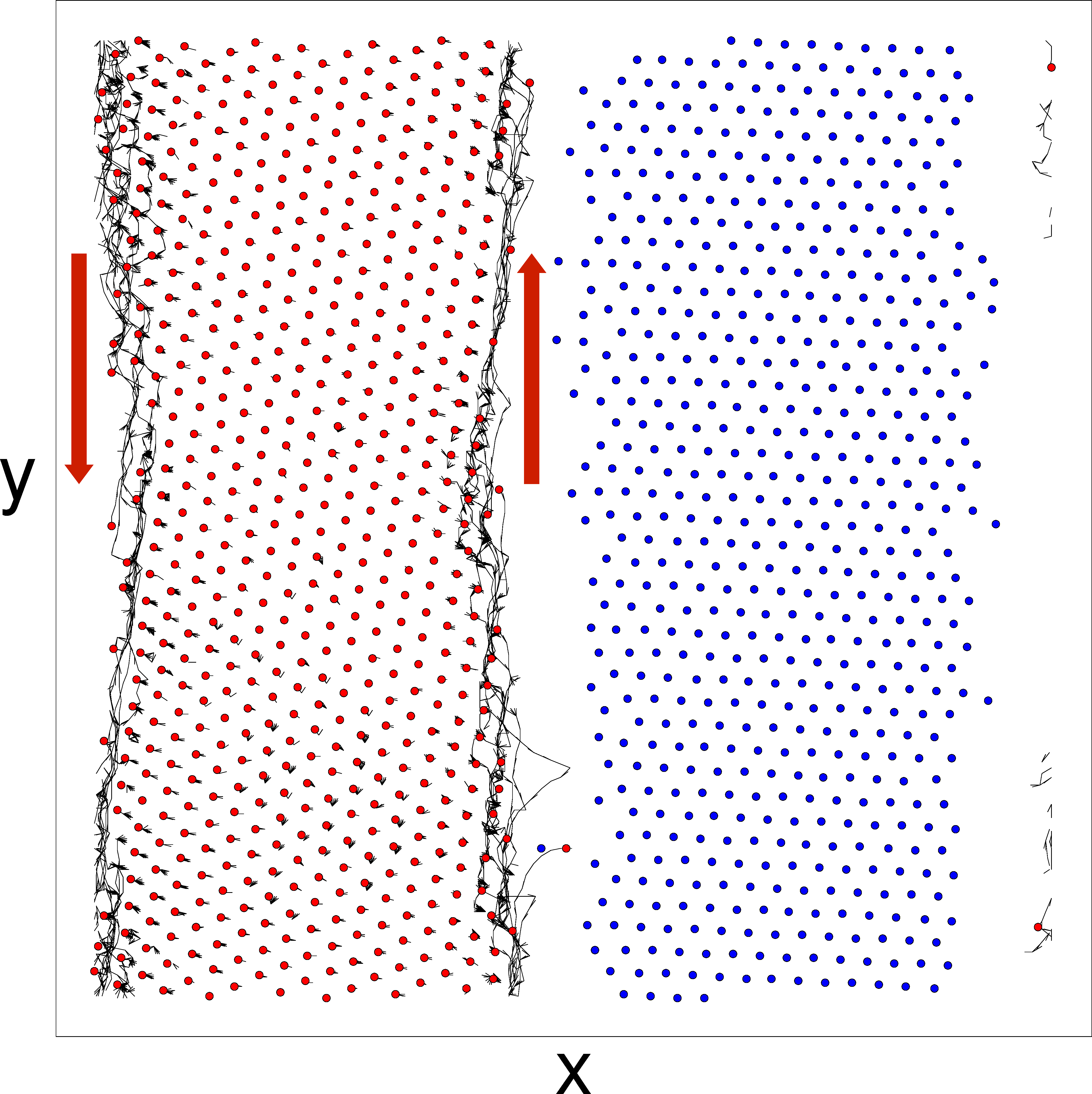}
\caption{Snapshot of disk positions for the active (blue) and
  passive (red) disks along with the trajectories of only
  the passive disks (lines)
  in a system with
  $A = 2.5$ and $\phi = 0.7272$.
  The motion in the
  bulk of the passive
  disks is reversible,
  but
  a net transport of disks
  occurs along the edges, as indicated
  by the trajectories.
  The arrows indicate
  the direction of the currents.
  For clarity, the disks are drawn at one-third their actual size.
}
\label{fig:8}
\end{figure}

In Fig.~\ref{fig:8} we illustrate the passive
disk trajectories
at $A = 2.5$
and $\phi = 0.7272$, where
the system is
completely phase separated.
In the bulk of the passive disk region, the disks
move in a reversible periodic fashion, while
along the edges the passive disks
undergo net transport.
On the right edge, the passive disks
move in the positive $y$ direction,
as indicated by the arrow,
while the passive disks on the left edge
move
in the negative $y$ direction,
giving a net current of zero.
The behavior of the active disks has a similar feature,
with active disks in the bulk undergoing collective periodic reversible
motion, while the active disks along the edges
exhibit directed transport. 
The appearance of edge currents for
active chiral systems has previously been observed
for active spinning particles 
consisting of mixtures of Janus particles where
one species orbits clockwise and the other orbits counterclockwise
\cite{30}.  In the Janus particle system, partial phase separation occurs
and currents appear along the phase boundaries.
Edge currents have also been found
in models with active spinners in confinement
\cite{34},
as well as in
other studies of active rotators
\cite{46}. There
are also studies of colloids undergoing oscillatory motion
on patterned substrates
where edge motion occurs along interfaces \cite{39}.

\begin{figure}
\includegraphics[width=\columnwidth]{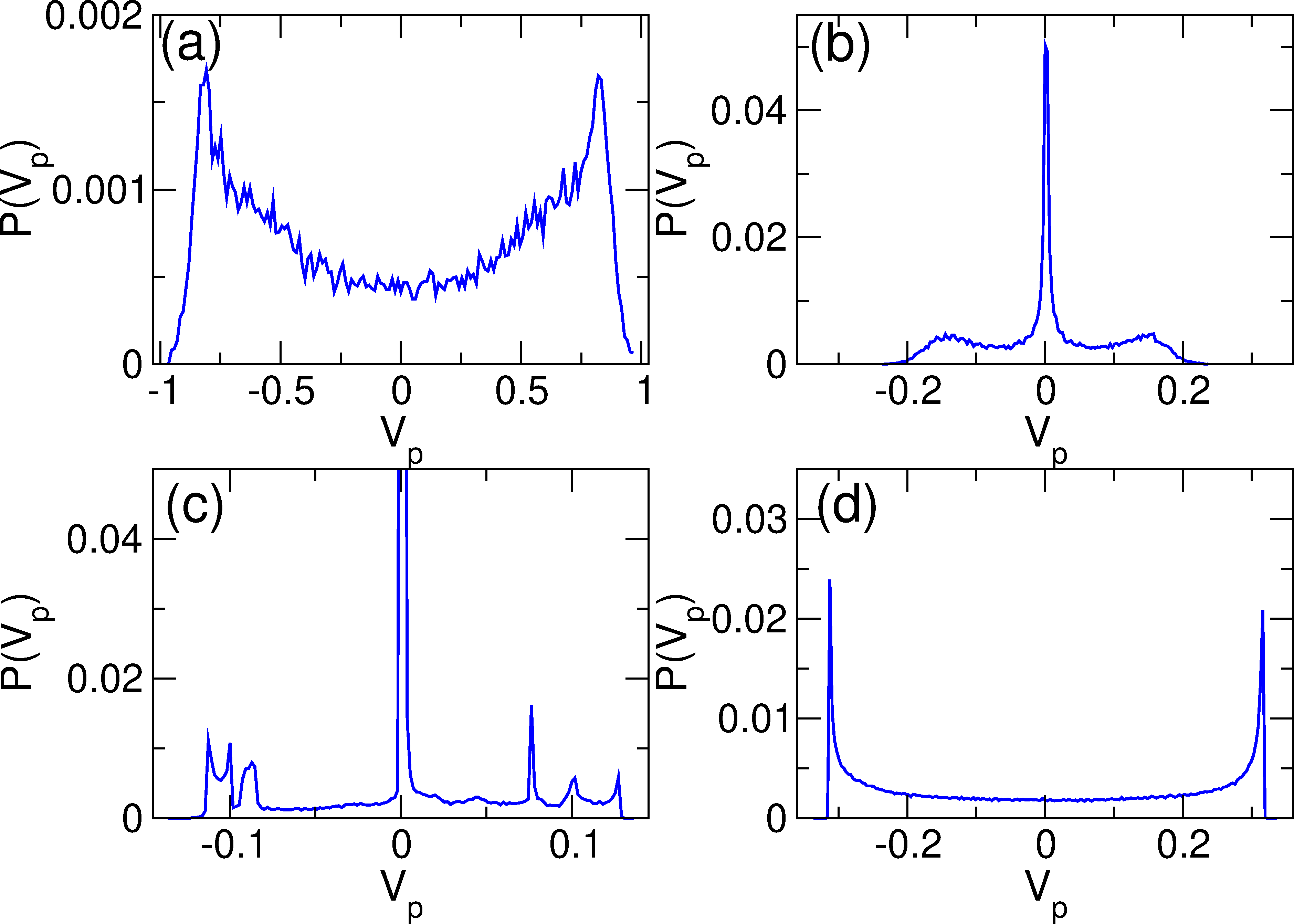}
\caption{$P(V_p)$,
  the distribution of the instantaneous velocities of the passive disks,
  after 600 circular drive cycles.
  (a) The fluctuating liquid state at $A = 5.5$ and $\phi=0.67$.
  (b) $A = 3.5$ and $\phi=0.67$, where
  edge transport occurs. 
  (c) $A = 2.5$ and $\phi=0.67$, in the reversible colliding phase.
  (d) $A=0.5$ and $\phi = 0.8847$, where the system
forms a reversible jammed state.  
}
\label{fig:9}
\end{figure}

We can characterize
the different phases
by analyzing the distribution of the instantaneous velocity $P(V_{p})$.
For phases in which no collisions occur,
$P(V_{p})$ exhibits a single peak at $V_{p} = 0.0$.
In Fig.~\ref{fig:9}(a) we 
plot $P(V_{p})$ after 600 circular drive cycles
for a system with $\phi = 0.67$ and $A = 5.5$ in the fluctuating liquid state. 
The peaks at negative and positive values of $V_p$ fall close to
the positions expected
for particles
undergoing sinusoidal motion,
and there is a local minimum at $V_{p} = 0$,
indicating that most of the passive disks are in 
continuous motion.
At $A=3.5$ in Fig.~\ref{fig:9}(b),
a state with edge transport forms and
there are two
peaks at finite $V_{p}$
that correspond to the passive disks moving along the edges as well as
a large peak at $V_{p} = 0.0$
produced by the stationary bulk passive disks.
In Fig.~\ref{fig:9}(c)
at $A = 2.5$,
the system
is in the reversible collision phase
and we find a large peak at $V_{p} = 0.0$
corresponding to the disks that are not colliding.
Multiple smaller peaks are present at finite values of $V_p$
due to the particular characteristics of
the reversible orbits that occur
during the
collisions.
In contrast, when there is edge transport,
the motion along the boundaries is irreversible, so there are only 
two smeared out peaks as shown in
Fig.~\ref{fig:9}(b).
In a sample with $\phi=0.8847$ and $A=0.5$, the system forms a reversible
jammed state in which all the particles move together rigidly,
and Fig.~\ref{fig:9}(d) indicates that
the shape of $P(V_p)$
matches what would be expected for completely sinusoidal motion.

\begin{figure}
\includegraphics[width=\columnwidth]{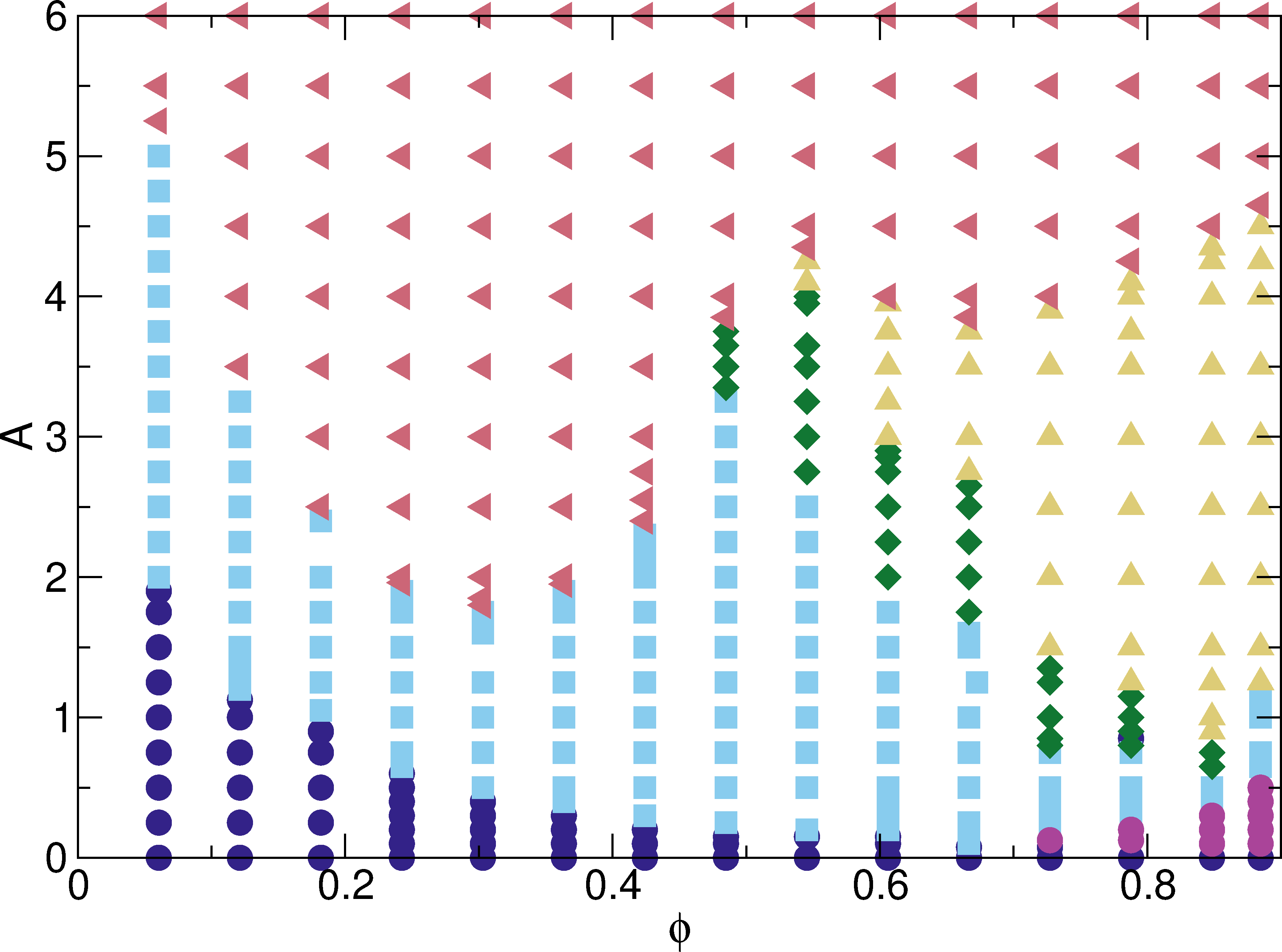}
  \caption{Dynamic phase diagram
    as a function of the circular drive amplitude
    $A$ vs $\phi$ for the binary system.
    Dark blue circles:
    a mixed reversible state with no relative motion of the particles.
    Light blue squares: a reversible pattern forming 
    or labyrinth state.
    Green diamonds: a completely phase separated state
    in which reversible collisions occur along the phase boundaries.
    Yellow triangles: a phase separated state that exhibits edge transport.
    Dark pink triangles:
    a fluctuating liquid state.
    Magenta circles: a reversible jammed state.    
}
\label{fig:10}
\end{figure}

By conducting a series of simulations for varied $\phi$ and $A$
and examining the behavior of $V_{p}$ and the images, we 
construct a dynamic phase diagram as a function of
$A$ vs $\phi$, as shown in Fig.~\ref{fig:10}.
At small $\phi$ and small $A$,
the system immediately enters
a non-interacting mixed state.
This state extends over the largest range of $A$ when $\phi$ is small,
since under these conditions
the distance between adjacent disks is 
large
and thus it is necessary for
$A$ to be relatively large in order for the disks to
contact each other.
In the region marked with light blue squares,
a reversible pattern forming or labyrinth state appears
which reaches its widest extent
at $\phi = 0.484$.
The diamonds indicate
the completely phase separated RC state in which the passive and
active particles undergo collisions but the system remains reversible.
The yellow triangles show the regime containing
phase separated states that undergo
edge transport, the dark pink triangles mark 
the disordered states, and
the magenta circles appearing at large $\phi$ show the presence of
what we call a reversible jammed state.

\begin{figure}
\includegraphics[width=\columnwidth]{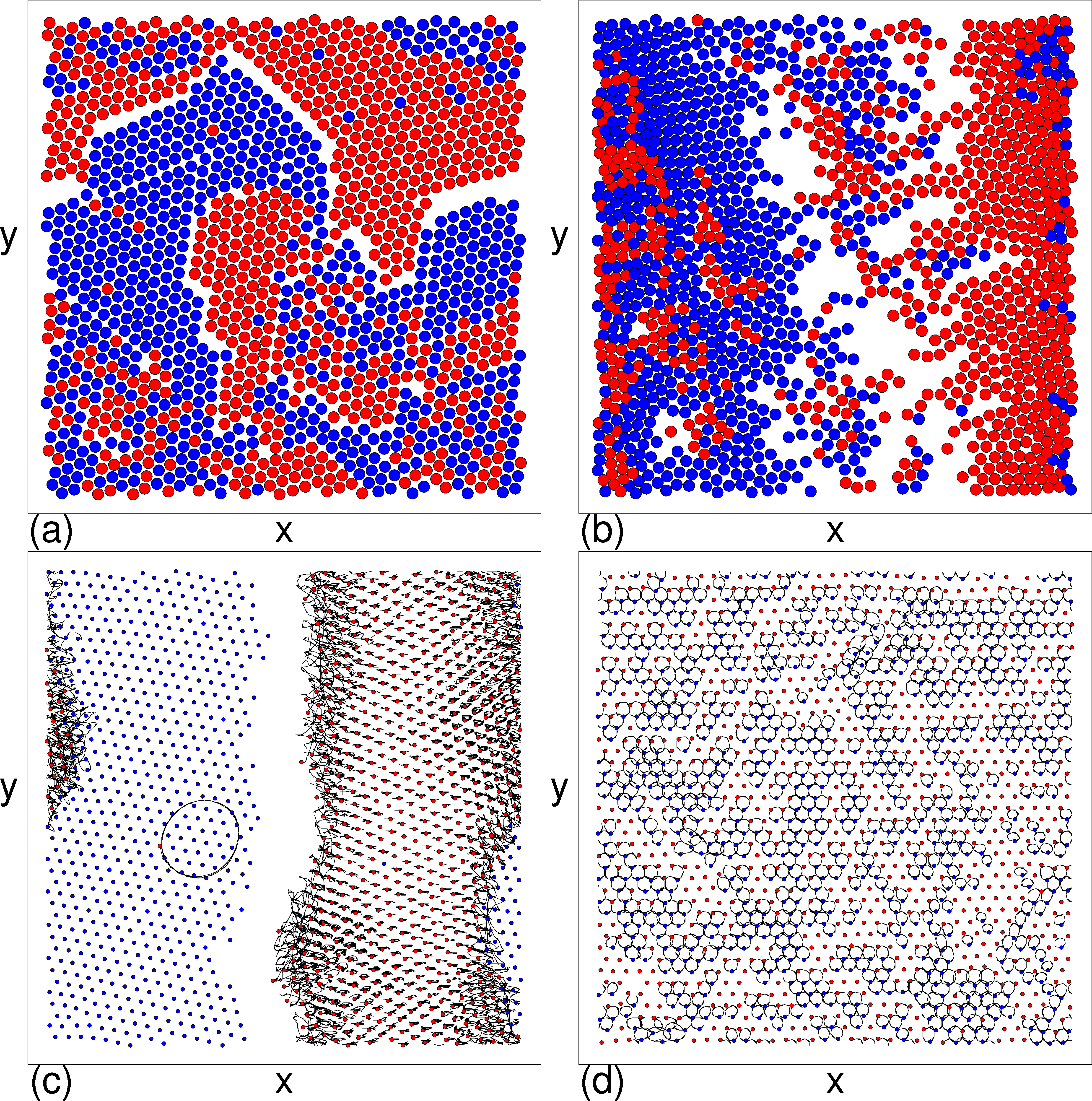}
\caption{
  (a, b) Snapshot of instantaneous positions for the active (blue) and passive
  (red) disks.
  (a) At $\phi = 0.848$ and $A = 0.5$, there is a
  pattern forming state
  in which the particles exhibit reversible collisions.
  (b) At $\phi = 0.848$ and $A = 5.0$,
  we find a fluctuating state where
  some local phase separation occurs.
  (c) Snapshot of disk positions and passive disk trajectories
  during five circular drive cycles for
  the active (blue) and passive (red) disks
  in the edge transport state at $\phi = 0.848$ and $A = 1.5$.
  For clarity, the disks are
  drawn at one-third their actual size.
  There is a single passive disk that has become
  trapped in the active disk cluster,
  and it rotates with the
  active cluster as indicated
  by the large circular orbit.     
  (d) Snapshot of disk positions and active disk
  trajectories for the active (blue) and passive (red) disks
  in the reversible jammed state for the system in Fig.~\ref{fig:9}(d)
  at $\phi=0.8847$ and $A=0.5$
  where there is no mixing relative to the initial configurations
  but
  the disks undergo
  reversible motion and the entire system moves as a rigid solid.
}
\label{fig:11}
\end{figure}

An example of a
pattern forming reversible phase that appears
for $\phi > 0.67$ is illustrated in
Fig.~\ref{fig:11}(a) at $\phi = 0.848$ and
$A = 0.5$.
In this state, collisions between the disks occur yet the motion
remains reversible.
We do not distinguish such states from the
pattern forming states that form without collisions between disks.
For $\phi < 0.484$, the disordered or liquid state
that appears at higher $A$ is generically strongly mixed, while
for $\phi > 0.484$,
the disordered phase
exhibits a partial phase separation of the type illustrated
in Fig.~\ref{fig:11}(b) 
at $\phi = 0.848$ and $A = 5.5$,
even as all of the
disks undergo diffusive behavior at long times.      
For $\phi > 0.67$,
there is a window in which the
disks remain in their initially diluted triangular
configuration
but undergo no mixing,
producing a state that can be viewed as reversibly jammed,
as illustrated
in Fig.~\ref{fig:9}(d) and
Fig.~\ref{fig:11}(d).
At higher $\phi$, near but below
the $\phi=0.9$ crystallization density,
a small but finite $A$ causes the effective radius of the
active disks to be larger than the actual radius,
so
the system
behaves as though it has reached the
crystallization density,
leading to the formation of the
reversible jammed state.
Once $A$ becomes
large enough, however,
plastic rearrangements occur and produce a partial phase separation.

The studies described here were performed in the absence of thermal
fluctuations.
It is possible that including thermal fluctuations would
permit
some portion
of the pattern forming phases
to become completely phase separated over time.
Thermal fluctuations would cause the disordered liquid phases
to expand,
and similarly the regions containing
edge currents could also expand.
We
note that
in studies
of chiral polar mixtures,
regions of ordered and disordered phases were observed
as functions of various parameters \cite{40}; however, several of the phases
we find for non-polar mixtures do not appear
for the polar chiral particles \cite{40}.
Active systems containing run and tumble
or driven diffusive active particles in which
there is a mixture of
active and passive particles or
multiple active species
with different mobilities
also exhibit
phase separated
and partially phase
separated states of the
different species \cite{47,48,49,50}. 

The evidence for the
existence of an absorbing phase transition
is strongest for $\phi < 0.5$ from the reversible collision-free state
to the fluctuating liquid state. 
We have not been able to identify
clear time scales for the transition from
the reversible pattern forming states to the
phase separated states.
This transition is more subtle
since the time scales in this case could be
related to coarsening dynamics rather than to
a potential absorbing phase transition.
We are also not able to determine
whether the transitions from the phase separated
states with or without edge transport
into the fluctuating liquid phase also show
a diverging time scale.
Such a measurement is complicated to perform
in states with edge transport,
since we often observe that
the width of the region in which the
irreversible edge current flows grows for higher $A$, 
as illustrated in Fig.~\ref{fig:11}(c)
for a sample with $\phi = 0.848$ and $A = 1.5$,
before completely disappearing 
at higher $A$.  

In Fig.~\ref{fig:11}(a)
we show the disk configurations
in the reversible pattern forming phase at
$\phi = 0.848$ and $A = 0.5$ where
large regions of triangular ordering
appear.
There are several active disks that have become trapped in the bulk region
of passive disks and are unable to move or escape.  Similarly, we
find
an occasional passive disk that is trapped in the bulk region of
active disks and rotates in synchrony with the active disks.
Figure~\ref{fig:11}(b) illustrates
the disordered or liquid phase
at $\phi = 0.848$ and $A = 5.0$, where
the system tends to phase separate but mixing of the disk species
still occurs,
and the passive disks undergo diffusive motion at long times.
In Fig.~\ref{fig:11}(c),
an edge transport state at $\phi = 0.848$ and $A = 1.5$
contains a
passive disk trapped in the bulk of the active disks that moves in a
circular orbit, while the remaining passive disks exhibit significant motion
only along the phase boundary between passive and active disks along with a
small circular motion in the bulk.
Figure~\ref{fig:11}(d) shows the trajectories
for the active particles at  $\phi = 0.8847$ and $A = 0.5$ 
from the system in Fig.~\ref{fig:9}(d) in 
a reversible jammed state where there is no phase
separation beyond the initial
configuration but the disks
are reversibly colliding and the entire system rotates as a rigid solid.

\begin{figure}
\includegraphics[width=\columnwidth]{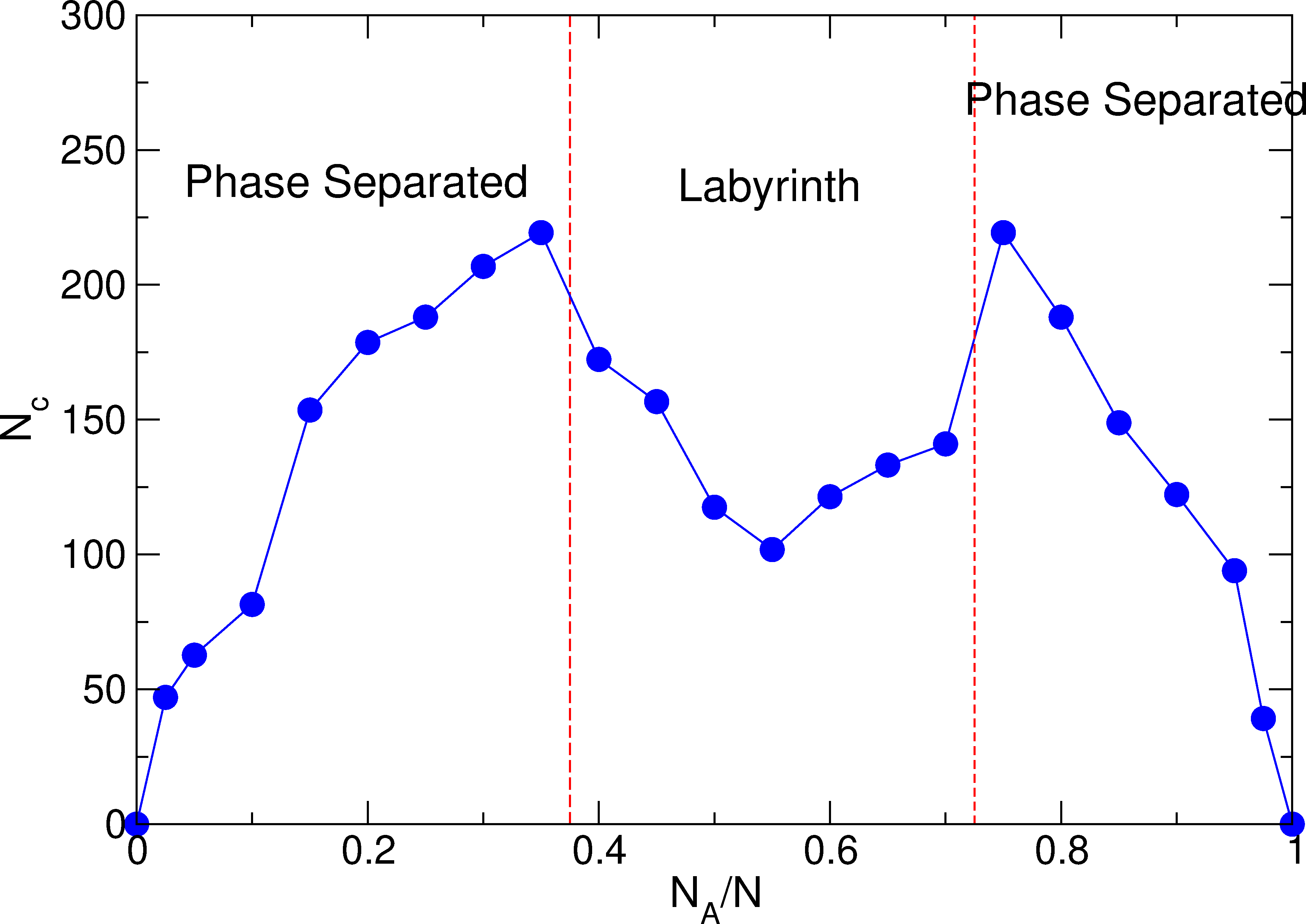}
\caption{$N_c$, the number of drive cycles required
  to reach a collision-free state, versus the fraction of
  active particles $N_A/N$
  for a system with $\phi = 0.424$ and $A = 2.0$.
  For $N_{A}/N \leq 0.35$,
  the system forms a stripe structure
  as illustrated in Fig.~\ref{fig:13}(c),
  while for $0.35 < N_{A}/N < 0.75$, a labyrinth state
  appears as shown in
  Fig.~\ref{fig:13}(a,b).
  For $N_{A}/N > 0.75$ the system
  enters another phase separated stripe state
  of the type shown in Fig.~\ref{fig:13}(d). }
\label{fig:12}
\end{figure}

\begin{figure}
\includegraphics[width=\columnwidth]{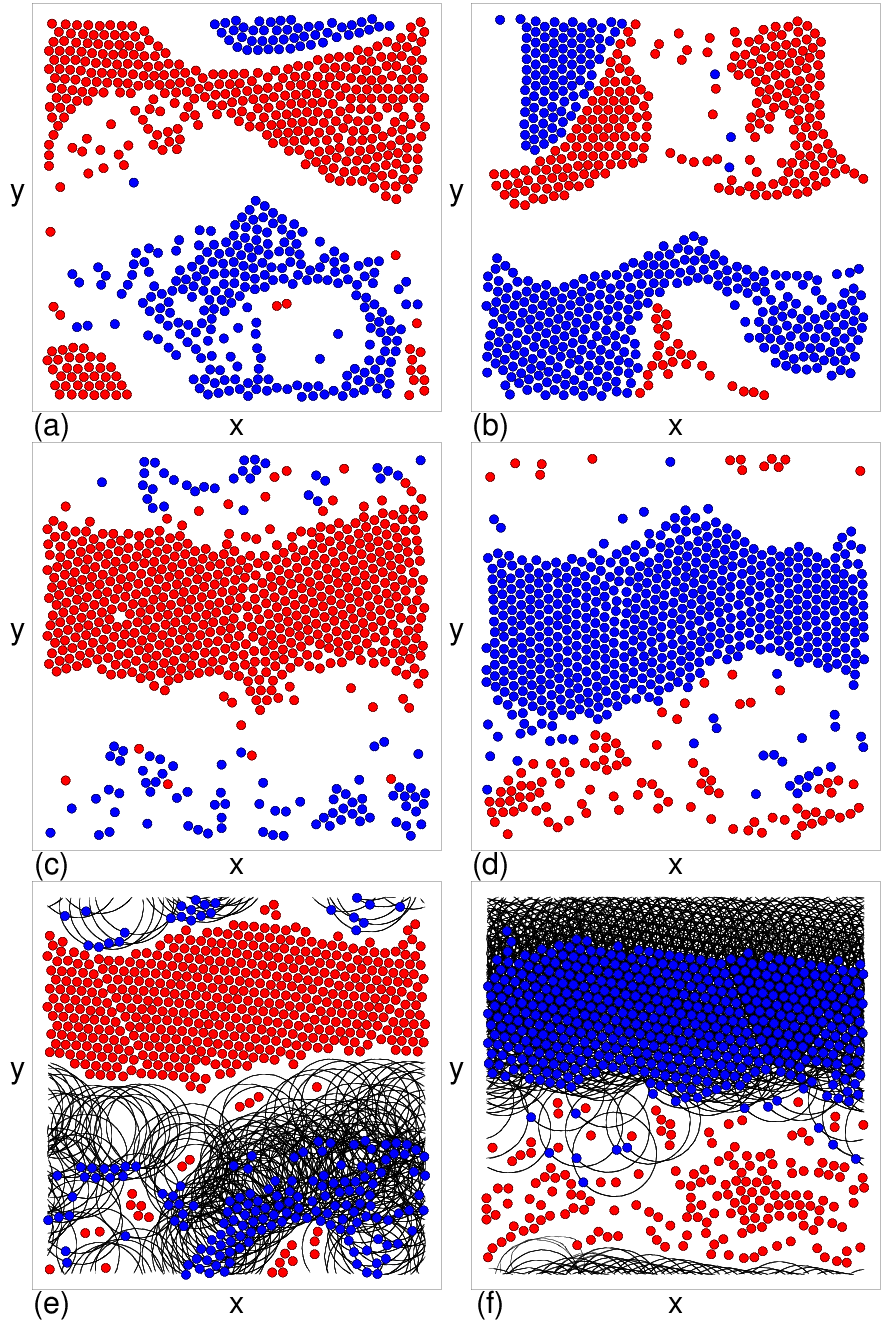}
\caption{
  (a-d) Snapshot of instantaneous positions for the active (blue) and
  passive (red) disks
  for the system in Fig.~\ref{fig:12}
  at $\phi = 0.424$ and $A = 2.0$.
  (a) $N_{A}/N = 0.4$.
  (b) $N_{A}/N = 0.6$.
  (c) $N_{A}/N = 0.15$.
  (d) $N_{A}/N = 0.85$.
  (e, f) Snapshot of disk positions and active disk trajectories for the
  active (blue) and passive (red) disks
  in the same system at (e) $N_{A}/N = 0.25$
  and (f) $N_{A}/N = 0.75$. 
}
\label{fig:13}
\end{figure}

We have also considered varied ratios of
active to passive disks and in general find similar results. 
In Fig.~\ref{fig:12} we plot $N_c$, the number of cycles required
to reach a reversible state in which all collisions are lost, versus the
fraction of active particles $N_{A}/N$ for a system in which
we fix the 
total disk density and drive
to $\phi = 0.424$ and $A = 2.0$, respectively.
We find that $N_c$ 
is small when $N_A/N<0.1$ or $N_A/N>0.9$
since most of the disks are already effectively phase separated into
regions containing only the majority species.
A local minimum in $N_c$ appears near $N_A/N=0.5$ in the labyrinth
state,
while there are
local maxima in $N_c$ near $N_A/N=0.35$ and
$N_A/N=0.75$ at the transitions between the phase separated and labyrinth
states.
At the $N_c$ minimum falling at half filling, the amount of demixing that
needs to occur is largest since there are equal numbers of each
disk species, but the magnitude of the fluctuations in the system is also
largest due to collisions between different disk species, and as a result
the system can rapidly explore phase space in order to reach a reversible
state.
We find a weak asymmetry around $N_A/N=0.5$ in which $N_c$ is slightly
higher for $0.5-\delta N_A/N$ compared to its value for $0.5 + \delta N_A/N$,
where $\delta N_A/N<0.1$.
This is a result of the reduced amount of fluctuations in the system when
the number of active disks is smaller than 50\% compared to when it is
higher than 50\%, and the system can more rapidly reach a reversible
state when the
fluctuations are larger.
Labyrinth patterns appear when $0.35 < N_{A}/N < 0.75$, as
indicated by the dashed lines in Fig.~\ref{fig:12},
while outside of this window strongly phase separated configurations
form.
In Fig.~\ref{fig:13}(a) we illustrate the collision-free state
for the system in Fig.~\ref{fig:12} at $N_{A}/N = 0.4$,
and in Fig.~\ref{fig:13}(b) we show the system at
$N_{A}/N = 0.6$.
In each case we find a labyrinth pattern,
and the smallest labyrinth
widths appear for
$N_{A}/N = 0.5$.  
The collision-free state at $N_A/N=0.15$ is shown in
Fig.~\ref{fig:13}(c),
where the passive disks
form a single dense cluster or stripe with
mostly triangular ordering while the active disks
form a non-clustered disordered arrangement.
At $N_A/N=0.85$ in
Fig.~\ref{fig:13}(d),
the active disks form a dense
cluster and the passive disks are in disordered positions.
The peaks in $N_c$
in Fig.~\ref{fig:12}  correspond to disk ratios for which
the
system first starts to form single domains
instead of labyrinths,
indicating that that there is a competition between the two different
types of pattern formation.  

\section{Other Types of Driving}

We next consider the case where $N_A=N/2$ and
both disk species are driven.
When all the disks are active with the 
same chirality, driving amplitude, and driving frequency,
there are no disk-disk collisions and the system rotates
in unison as a whole. 
If the second disk species moves counterclockwise
while the first disk species moves clockwise,
we observe the same behavior found for mixtures of active and passive
disks, except that the values of $A$ and $\phi$ at which
the reversible to irreversible
transition appears shift downward.
If the second species is subjected to a one-dimensional
$y$-direction ac drive
instead
of a circular drive, while the first species continues to undergo
circular motion,
we observe several different phases depending on whether
the ac drive is in or out of phase with the circular drive.

\begin{figure}
\includegraphics[width=\columnwidth]{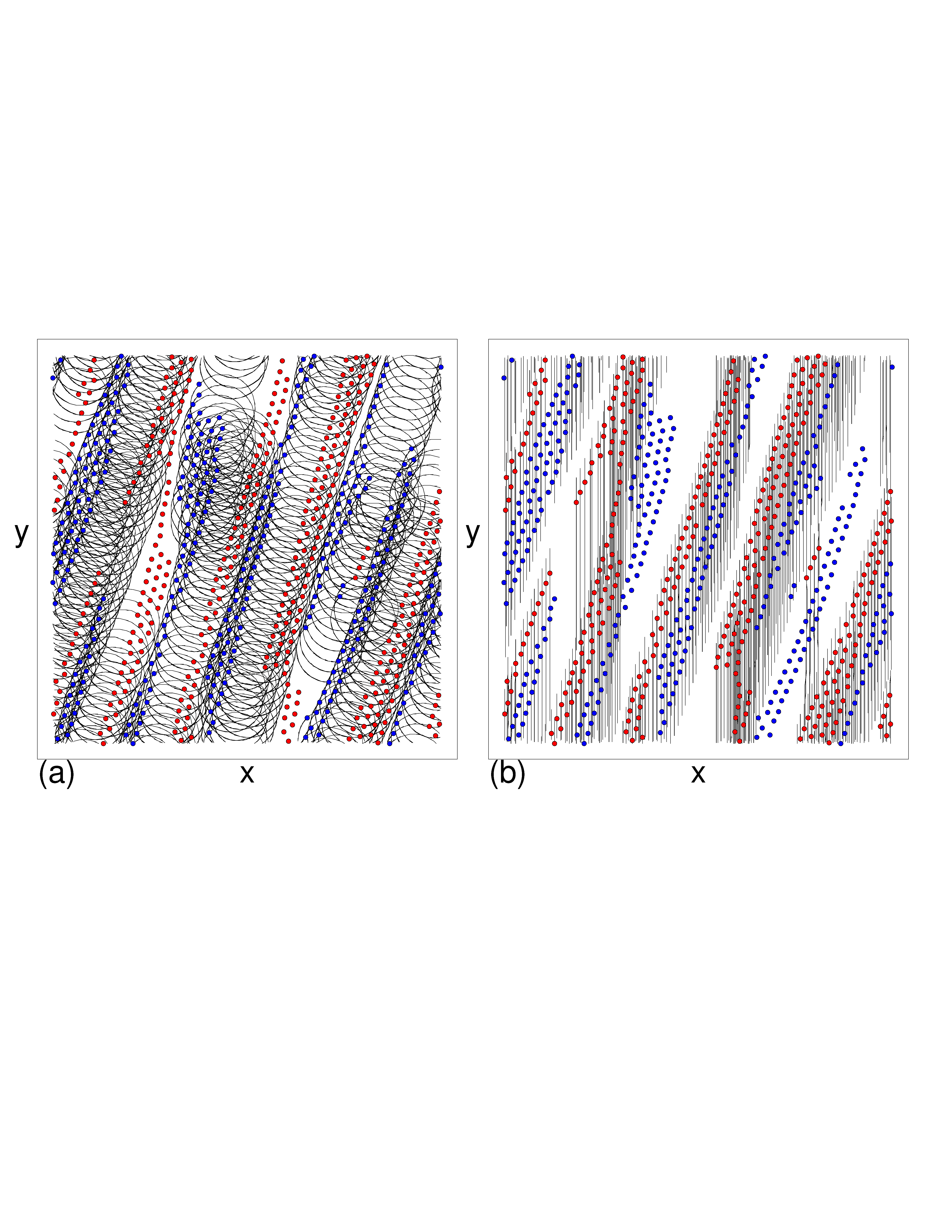}
\caption{Snapshot of disk positions and trajectories for
  species one (blue) and species two (red) disks in a system
  with $\phi=0.4242$
  where
  species one undergoes clockwise circular motion with $A=1.5$ and
  species two undergoes $y$-direction ac driving with
  $A_{2} = 4.0$.
  For clarity, the disks are drawn at one-half their actual size.
  The drives of the two species are in phase.
  (a) Trajectories of only species one showing circular orbits. (b)
  Trajectories of only species two showing 1D motion.
}
\label{fig:14}
\end{figure}

\begin{figure}
\includegraphics[width=\columnwidth]{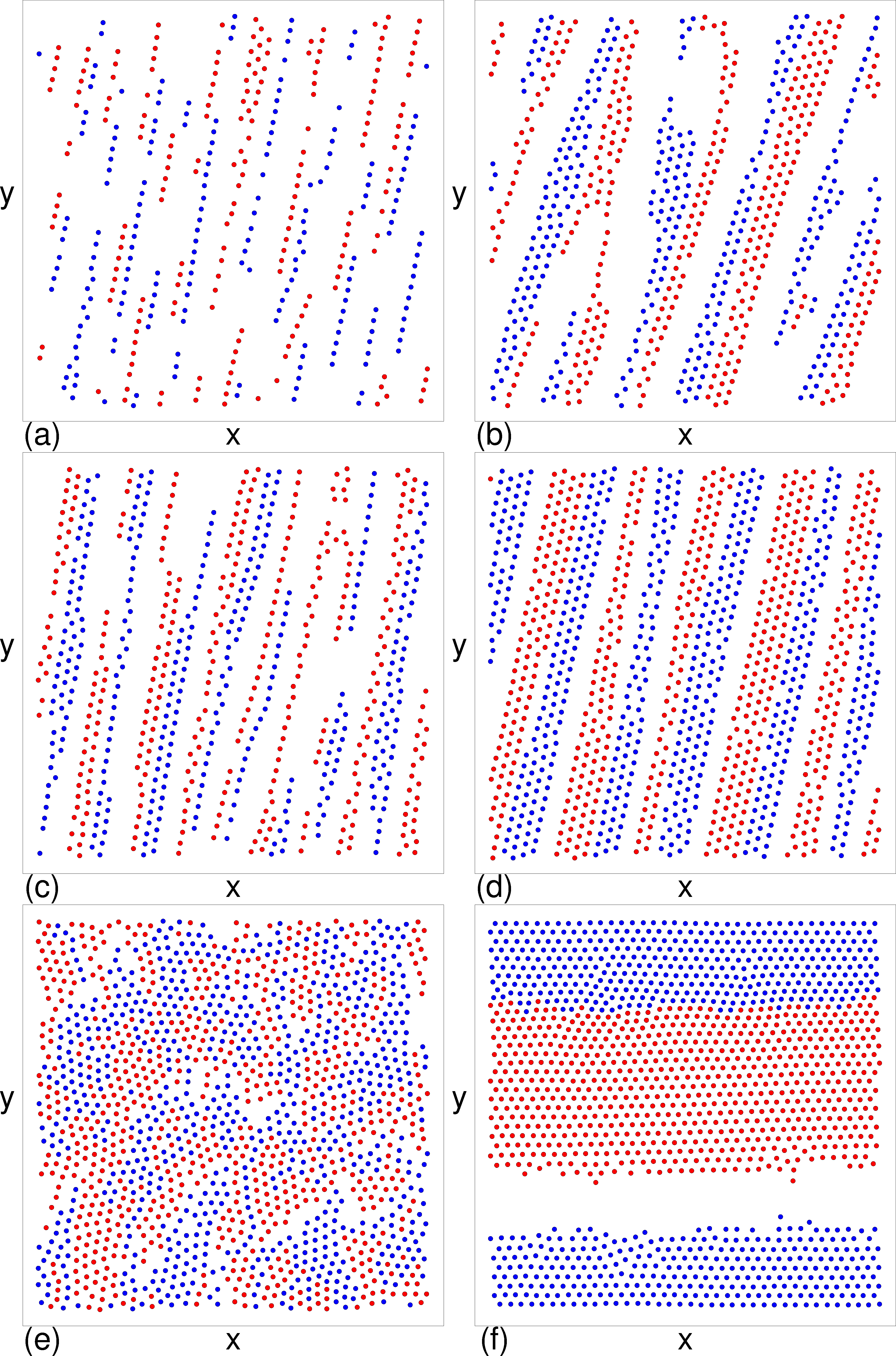}
\caption{Snapshot of instantaneous positions for the
  circularly driven species one disks (blue) and
  ac driven species two disks (red) for a system
  with $A=1.5$ in which the
  two species move in phase.
  For clarity, the disks are drawn at one-half their actual size.
  (a) Reversible laned state at $A_{2} = 7.0$ and $\phi = 0.182$.
  (b) Reversible laned state at $A_2 =4.0$ and $\phi = 0.4242$, which is
  also illustrated in 
  Fig.~\ref{fig:14}.    
  (c) Reversible laned state at $A_2=7.0$ and $\phi=0.4242$.
  (d) Laned state at $A_2=7.0$ and $\phi = 0.67$.
  (e) Fluctuating liquid phase at
  $A_2=7.0$ and $\phi = 0.848$.
  (f) A completely phase
  separated state
  at $A_{2} = 1.5$ and  
  $\phi = 0.848$.
}
\label{fig:15}
\end{figure} 

We first consider the case in which disk species two experiences a
$y$-direction ac drive that is in phase with the circular ac drive
of species one, such that
$F^{y}_2 = A_{2}\cos(\omega t){\bf {\hat y}}$.
Here $\omega$ has the same value for both disk species.
In Fig.~\ref{fig:14}(a) we plot the positions of all the disks and the
trajectories of only the species one circularly driven disks for a sample
with
$\phi = 0.4242$, $A = 1.5$, and $A_{2} = 4.0$.
Figure~\ref{fig:14}(b) shows the same state with only the trajectories of
the species two ac driven disks.
At this value of $A$
we observe a new vertical stripe or laned phase
that is absent when the species two disks are passive.
As we vary $\phi$,
we also observe a fluctuating disordered state and phase separated states.
In Fig.~\ref{fig:15}(a) we plot 
the disk positions
in a sample with
$\phi = 0.182$, $A = 1.5$, and $A_{2} =  7.0$,
which is initially in
a disordered state but 
organizes over time into a
collision-free
reversible laned state
in which the particles are 
aligned along the $y$-direction.
In Fig.~\ref{fig:15}(b)
at $\phi = 0.424$ and $A_{2} = 4.0$,
which is also illustrated
in Fig.~\ref{fig:14},
the lanes are thicker and
there is a tendency for the two species to cluster together.
At $A_2=7.0$ in
Fig.~\ref{fig:15}(c),
the width of the
lanes is reduced,
while in
Fig.~\ref{fig:15}(d)
at $\phi = 0.67$ and $A_{2} = 7.0$,
the lanes are wider.
In general we find that
as $A_2$ increases, the lanes
become thinner, while the lanes get thicker
as $\phi$ increases.
The laned states for $\phi < 0.5$ are collision-free,
while for $\phi > 0.5$,
reversible disk-disk collisions can occur.
In Fig.~\ref{fig:15}(e)
at $\phi = 0.848$ and $A_{2} = 7.0$
we find a fluctuating liquid state
without lanes, while
in Fig.~\ref{fig:15}(f)
at $\phi = 0.848$ at $A_{2} = 1.5$,
a fully phase separated state appears.
In general, when the drives of the
two species are in phase,
the edge transport that occurs in the
fully phase separated states is weaker compared to the edge transport
exhibited by passive species two disks.

\begin{figure}
\includegraphics[width=\columnwidth]{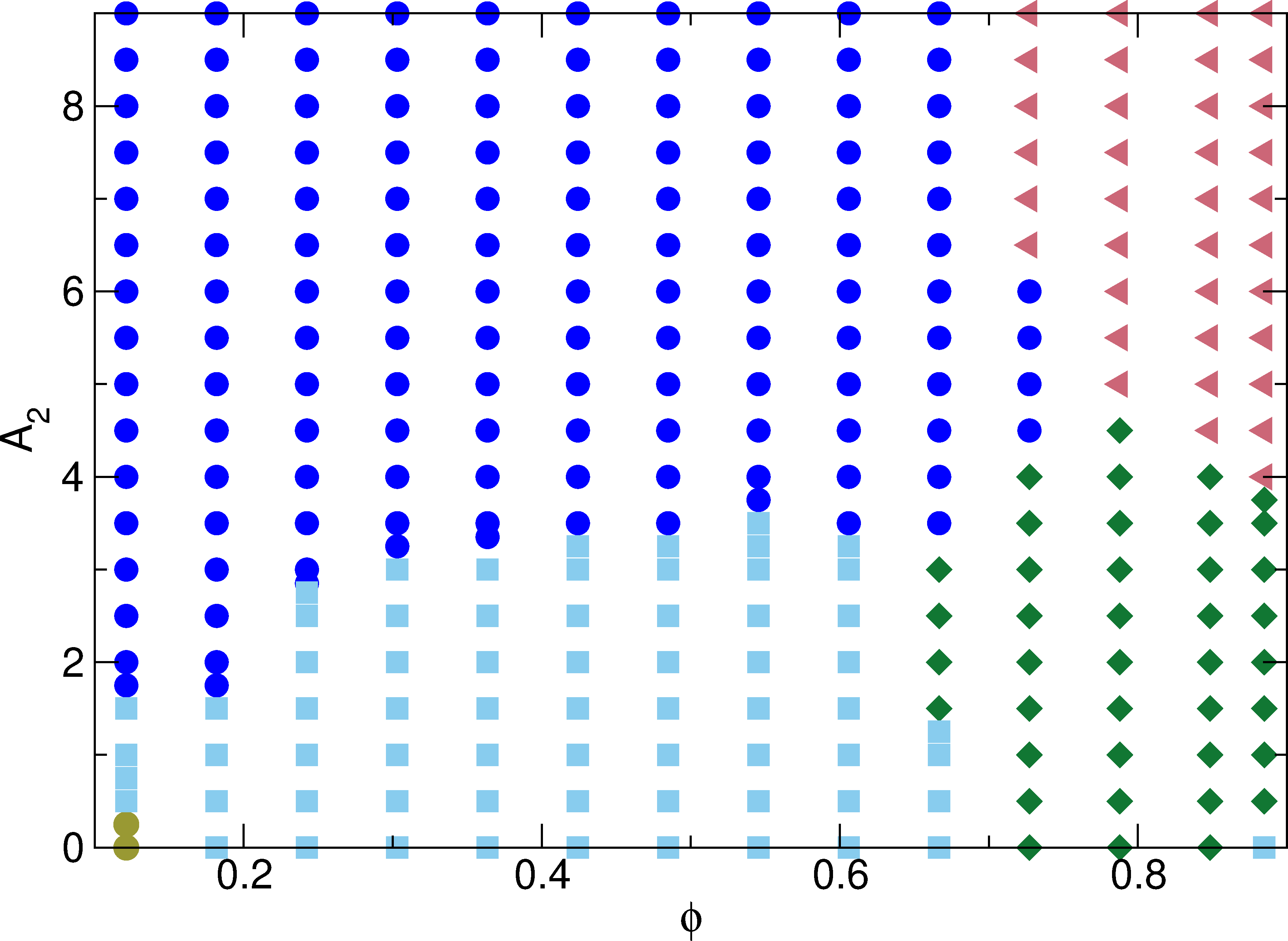}
  \caption{Dynamic phase diagram as a function of $A_{2}$ vs $\phi$
    for a system in which
    species one disks undergo
    clockwise circular motion 
    with $A = 1.5$
    and species two disks
    are subjected to a $y$-direction
    ac drive of amplitude $A_{2}$.
    Dark blue circles: reversible laned state.
    Light blue squares: reversible pattern forming state.
    Green diamonds: completely phase separated state.
    Dark pink triangles: fluctuating liquid state.
    Light green circles: initial reversible state.
    Here we do not distinguish
    the fully phase
    separated states that exhibit edge transport from those that do
    not exhibit edge transport. 
}
\label{fig:16}
\end{figure}

\begin{figure}
\includegraphics[width=\columnwidth]{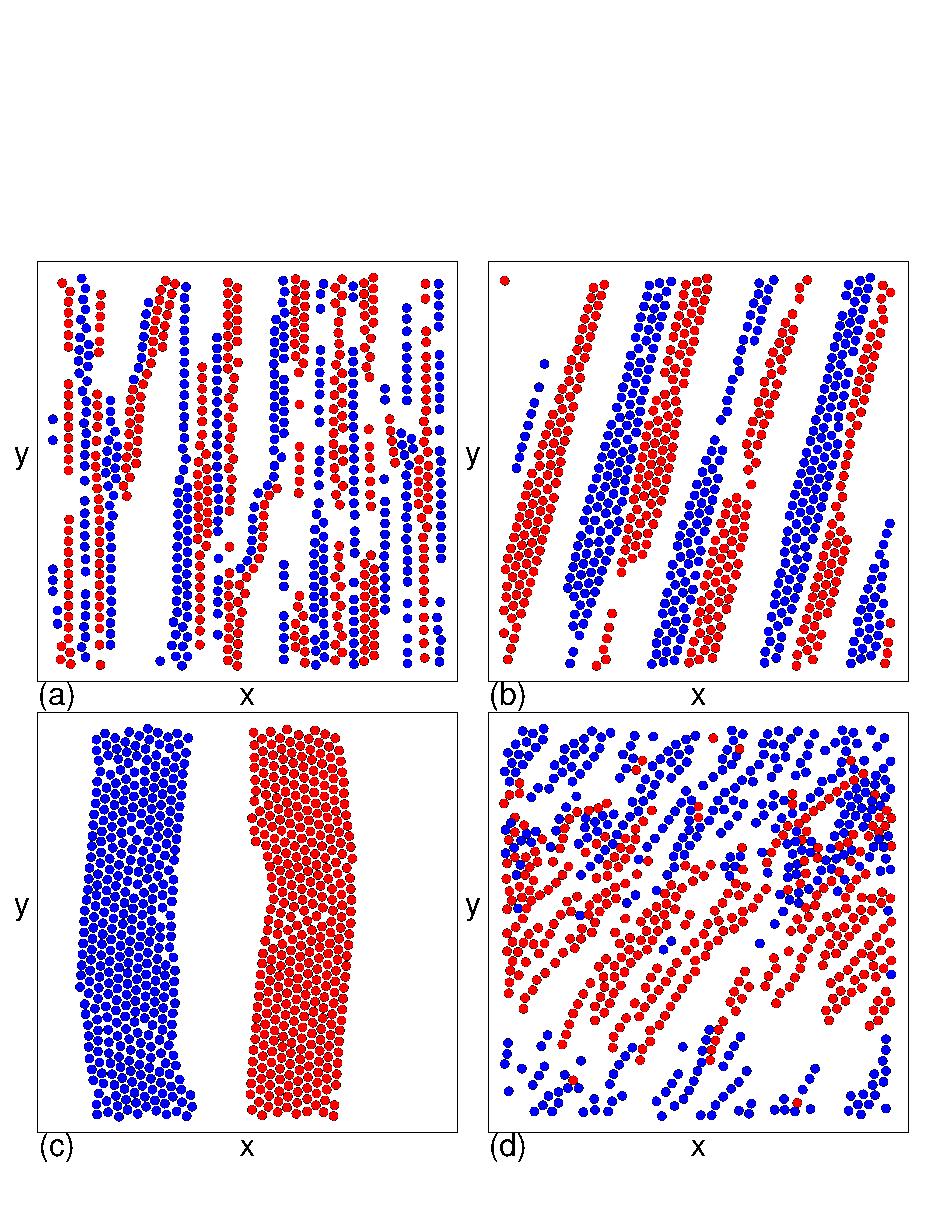}
\caption{
  Snapshot of instantaneous positions for the circularly driven species
  one disks (blue) and ac driven species two disks (red)
  for systems with $\phi = 0.4242$ and $A_2=7.0$.
  (a) $A = 0.0$.
  (b) $A = 2.5$.
  (c) $A = 4.5$.
  (d) $A = 8.0$.
}
\label{fig:17}
\end{figure}

In Fig.~\ref{fig:16} we plot a dynamic phase diagram
as a function of $A_{2}$ versus $\phi$ for the system
in Figs.~\ref{fig:13} and \ref{fig:14} 
in which
species one undergoes circular motion with
$A = 1.5$. 
We find a laned state, a pattern forming state, a phase separated state,
a fluctuating phase, and the initial reversible state in which the
particles do not rearrange.
Here we do not distinguish 
phase separated states that have edge transport from those that do
not have edge transport.
We have also
varied $A$ in addition to varying
$A_{2}$
and observe phases similar to those illustrated in Fig.~\ref{fig:16},
but find in general that the
fully phase separated state
remains stable for much lower values of $\phi$ compared to systems in which
the species two disks are passive. 
In Fig.~\ref{fig:17}(a) we plot the disk positions
for a system with $\phi = 0.4242$ at 
$A = 0.0$ and $A_{2} = 7.0$.
A laned state appears in which the lanes are aligned in the $y$ direction,
parallel to the ac drive on the species two disks.  This is distinct from
the laned state that forms at finite $A$ in which the lanes are
tilted, as illustrated in 
Fig.~\ref{fig:17}(b)
for $A = 2.5$ and $A_{2} = 7.0$.
In Fig.~\ref{fig:17}(c)
at $A = 4.5$ an $A_{2} = 7.0$,
complete phase separation occurs,
while
in Fig.~\ref{fig:17}(d)
at $A= 8.0$ and $A_{2} = 7.0$
we find a fluctuating liquid state.

\begin{figure}
\includegraphics[width=\columnwidth]{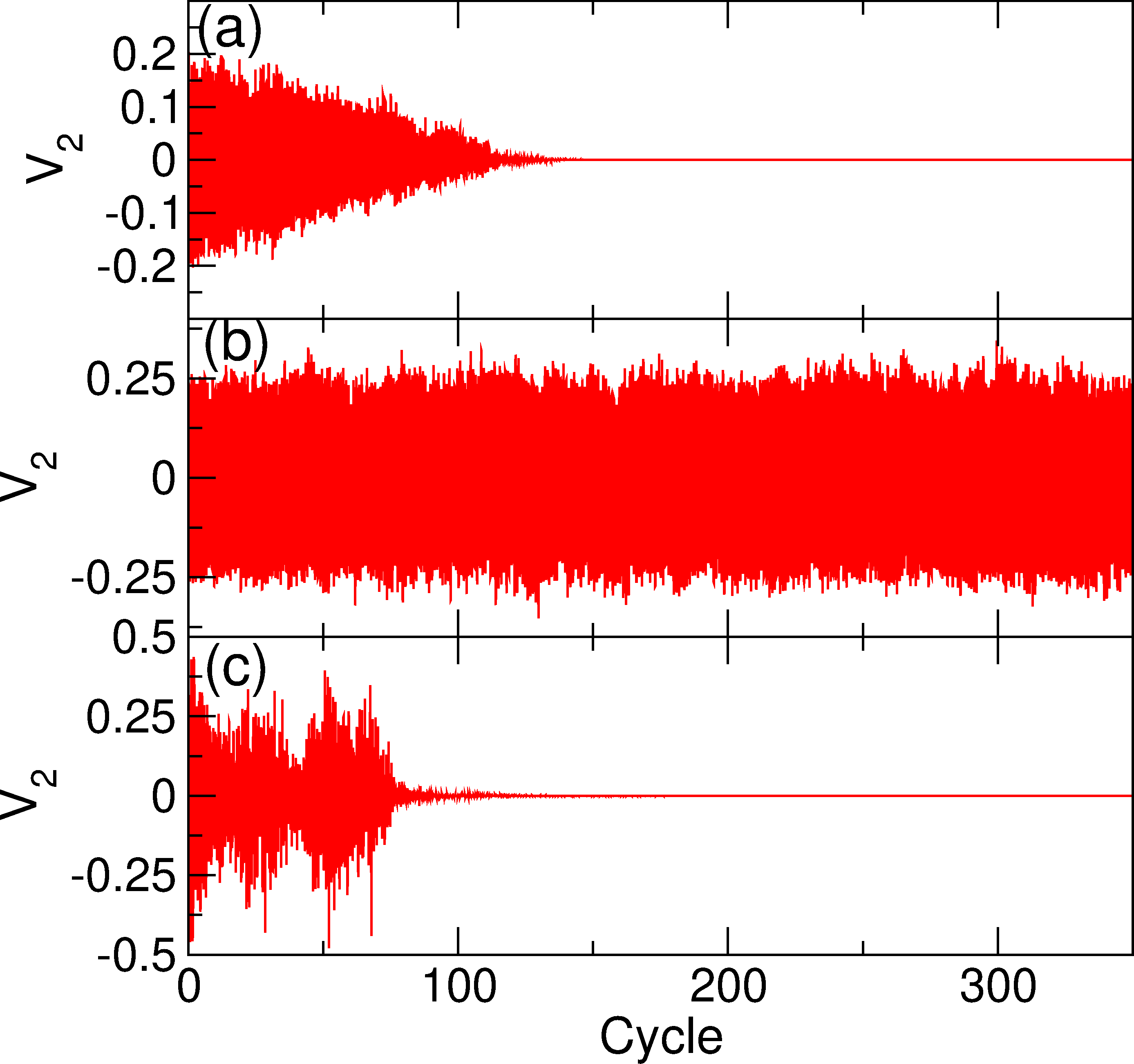}
\caption{
  $V_2$, the instantaneous $x$-component velocity of the species two disks,
  vs time in ac drive cycles
  for a system in which
  the species one disks are circularly driven
  and the species two disks experience a $y$-direction ac drive
  that is out of phase with the species one drive.
  Here  $\phi = 0.242$ and $A = 1.5$.
  (a) At $A_{2} = 0.5$, the system organizes
  into a collision-free pattern forming state.
  (b) At $A_{2} = 4.0$, the system remains in a fluctuating state.
  (c) At $A_{2} = 8.0$, the system forms another
  reversible collision-free state.    
}
\label{fig:18}
\end{figure}

\begin{figure}
\includegraphics[width=\columnwidth]{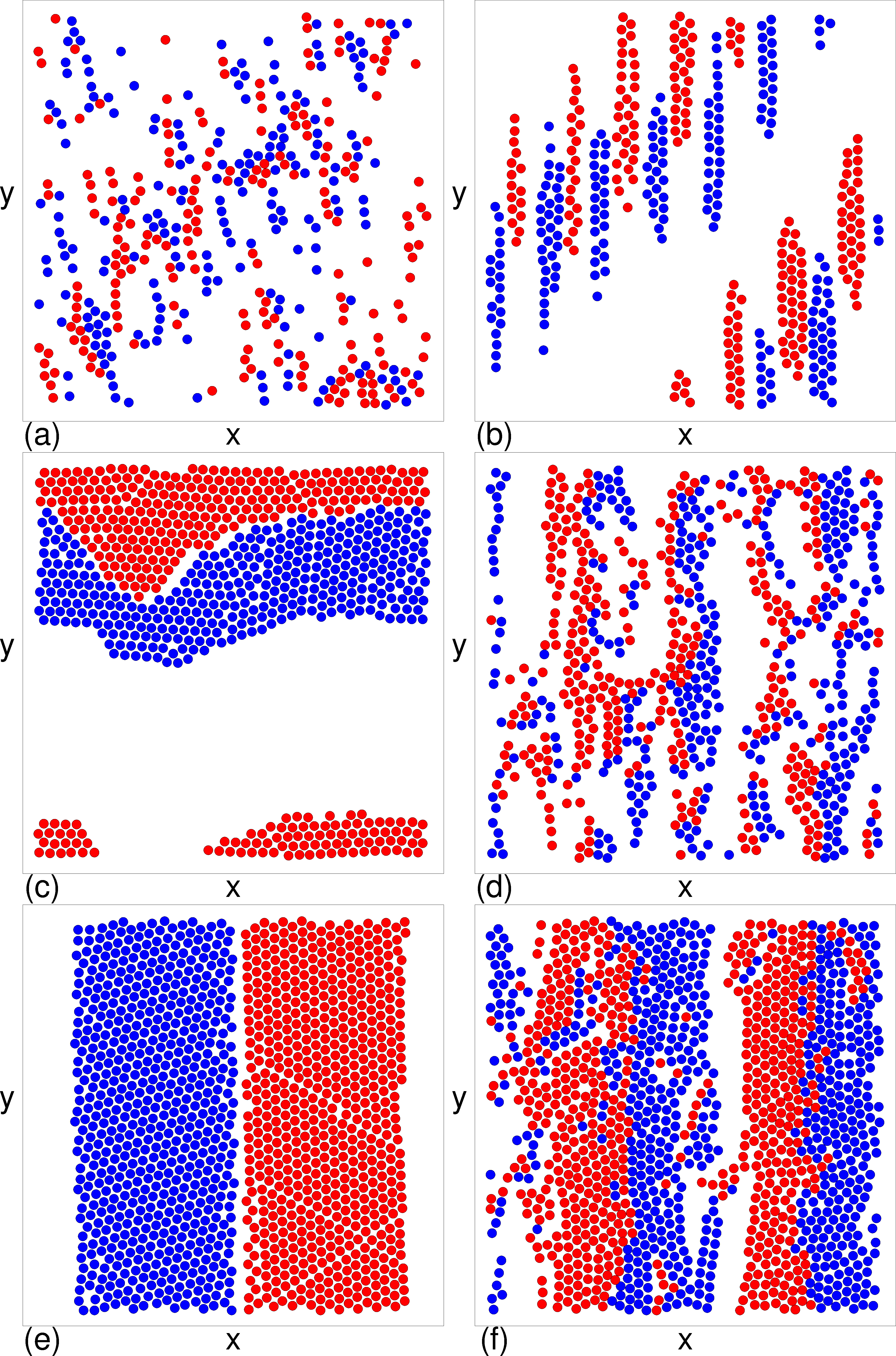}
\caption{
  Snapshot of instantaneous positions for the circularly driven species one
  disks (blue) and ac driven species two disks (red) in
  the system from Fig.~\ref{fig:18} where the
  two species are moving out of phase.
  (a) Irreversible state at $\phi=0.242$ and $A_2=4.0$.
  (b) Reversible mixed laned state at $\phi = 0.242$
  and $A_2=8.0$.
  (c) Phase separated state at $\phi=0.424$ and $A_2=3.5$.
  (d) Fluctuating state at $\phi=0.424$ and $A_2=8.0$.
  (e) Phase separated state with edge transport
  at $\phi=0.848$ and $A_2=4.5$.
  (f) Partially phase separated fluctuating liquid at
  $\phi=0.848$ and $A_2=8.0$.
}
\label{fig:19}
\end{figure}

We next consider
samples in which the ac drive on species two is out of phase with
the circular drive of species one, so that
$F^y_{2} = A_{2}\sin(\omega t){\bf {\hat y}}$.
Placing the two drives out of phase
tends to increase the frequency of collisions
between the particles.
It also produces the interesting effect of several reentrant phases, such
as reversible behavior
for low and high values of $A_{2}$ but
irreversible behavior at
intermediate values of $A_{2}$.
An example of this appears
in Fig.~\ref{fig:18}(a)
where we plot $V_2$, the instantaneous $x$ component velocity of
the species two disks,
versus time in number of ac drive cycles
for a system with $\phi = 0.242$,  $A = 1.5$, and $A_{2} = 0.5$.
Initially the disks are in a fluctuating state, but after 110 cycles, they
settle into a collision-free reversible state with
$V_{2} = 0.0$.
We note that the   
instantaneous $y$ component
velocity of the species two disks (not shown)
has strong periodic oscillations due to the ac drive.
For this value of $A_{2}$, the reversible state is
a pattern forming state similar to those previously shown.
In the same system at
$A_{2} = 4.0$,
Fig.~\ref{fig:18}(b) shows
that the sample remains in a fluctuating state as
illustrated
in Fig.~\ref{fig:19}(a), 
while at $A_2=8.0$ in
Fig.~\ref{fig:18}(c),
the system
settles into a collision-free reversible state.
The reversible state at $A_{2} = 8.0$
appears in Fig.~\ref{fig:19}(b).
It differs
from
the
pattern forming state found
at lower $A_{2}$
since it is composed of a series of
single-species vertical stripe segments of alternating species
type, and the stripe segments themselves form
a higher order diagonal stripe.
We label this phase
a mixed laned state. 
In Fig.~\ref{fig:19}(c) we show the
disk configurations at $\phi = 0.4242$ and $A_{2} = 3.5$
where the system forms a phase separated state, while
in Fig.~\ref{fig:19}(d)
at the same density and $A_{2} = 8.0$,
a fluctuating liquid phase appears.
Figure~\ref{fig:19}(e) shows a phase separated state
with edge transport at $\phi = 0.848$ and $A_{2} = 4.5$,
while in Fig.~\ref{fig:19}(e), a sample with the same density
and
$A_{2} = 8.0$
is in a liquid
state that has undergone partial phase separation.

\begin{figure}
\includegraphics[width=\columnwidth]{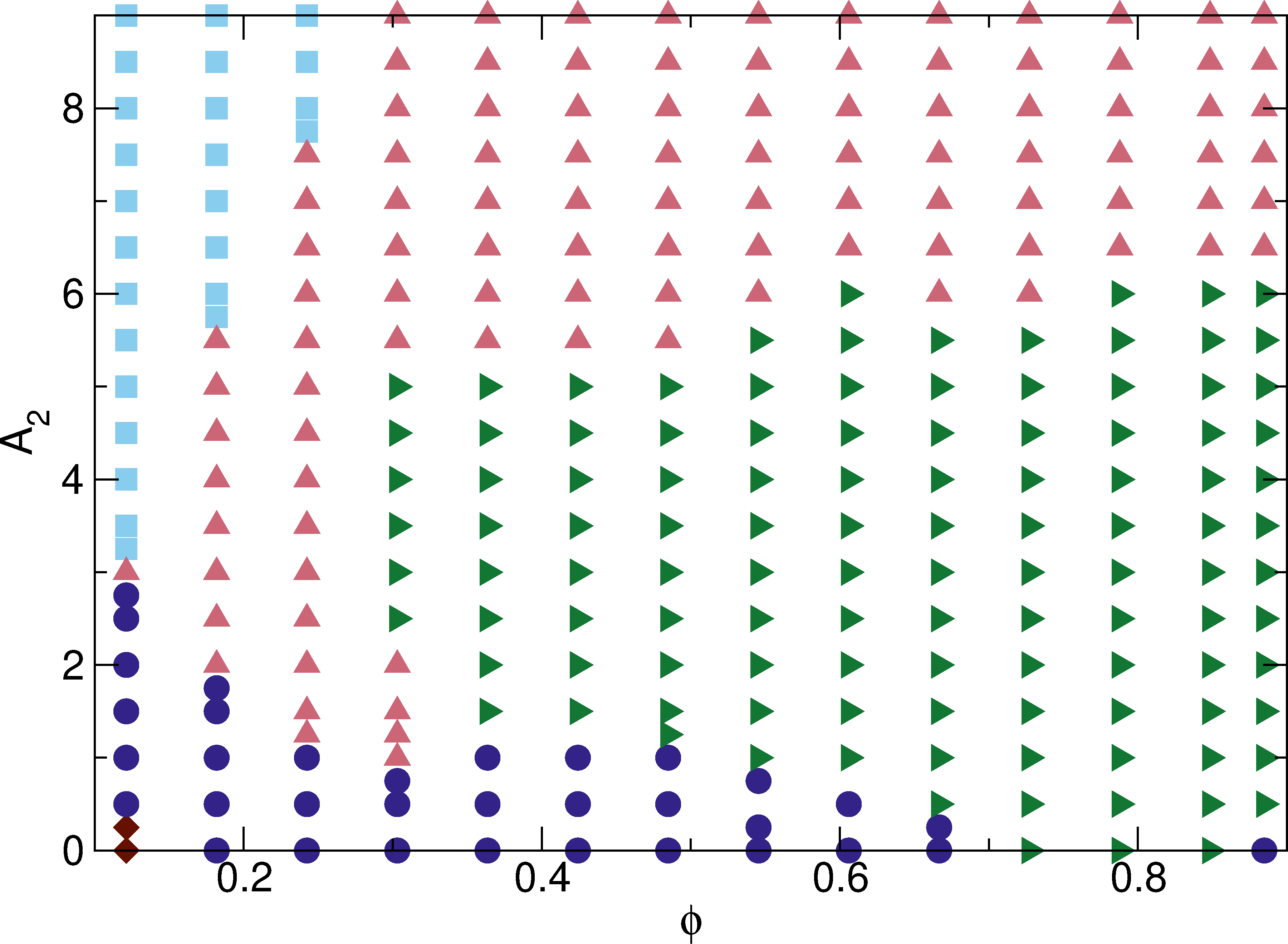}
  \caption{
    Dynamic phase diagram as a function of $A_2$ vs $\phi$ for the system
    in Figs.~\ref{fig:18} and \ref{fig:19} in which species one is
    circularly driven with $A=1.5$
    and species two has a $y$-direction ac drive that is
    out of phase with the species one motion.
    Dark blue circles: initially reversible state.
    Light blue squares: reversible pattern forming state.
    Green triangles: fully phase separated state.
    Dark red diamonds: mixed laning state.
    Dark pink triangles: fluctuating liquid state.
    We note that the fully phase
    separated states at the higher values of $A_{2}$ and
$\phi$ have edge transport. 
}
\label{fig:20}
\end{figure}

In Fig.~\ref{fig:20} we construct a dynamic
phase digram as a function of $A_2$ versus $\phi$
for the system in Figs.~\ref{fig:18} and \ref{fig:19}
at fixed $A = 1.5$.
We find
a reversible initial state,
a pattern forming state,
fully phase separated states
that can be completely reversible or have edge transport,
a fluctuating state,
and a reversible mixed laning state.
A reentrant phenomena appears for $\phi < 0.5$
when
the low $A_{2}$ reversible pattern forming state
is followed by
an intermediate irreversible 
state that transitions
into a reversible state of either the
mixed lane or phase separated type.
At $\phi  = 0.303$,
the system transitions from
a reversible pattern forming state to a fluctuating state,
than into a phase separated state, and
finally back to 
an irreversible  fluctuating state
as a function of increasing $A_{2}$.
We generally find larger regions of fluctuating states when the
drives of the two species are out of phase compared to when they
are in phase.
We have also examined several cases where we vary
both $A$ and  $A_{2}$ and find similar phases, including 
complete phase separation 
for $\phi < 0.5$ in which
stripes form that are parallel to the $x$ direction,
a configuration that is distinct from the $y$ direction stripes that
form when the drives of the two species are in phase.

\section{Summary} 
We examine a binary assembly of two-dimensional disks 
at an overall disk density $\phi$ in which
one disk species undergoes
chiral motion under a clockwise circular drive while the
second disk species is passive.
We find a rich variety of distinct
pattern forming and dynamical states.
As we vary the disk density and the radius of the active disk orbit,
the system is generally
initially 
in a fluctuating state where collisions between the disk
species occur, and then it gradually organizes
into either a reversible state or a
steady fluctuating state.
At low disk densities,
there are no disk collisions in 
the reversible state, 
and the system
forms labyrinth patterns
that become
coarser as the active orbit radius increases.
We
identify a critical orbit radius
at which the time required for the system to reach a collision-free
reversible state diverges as a power law with exponent
$\nu = -1.236$, a value that is
close to what is expected
for a conserved directed percolation transition.
This behavior
is similar to that
found in
other periodically driven  
systems that
organize to reversible states.
For higher disk densities,
the system
forms completely phase separated states 
that
can either be reversible
or exhibit edge currents or directed transport along the
boundaries between the two phases.
The direction of the edge current
is controlled by the chirality of the active disks.
We
identify several reversible 
phases in which
collisions or contact between the disks continues to occur, and
show that these can be
pattern forming states or a reversible jammed state.
For the case
in which one disk species
undergoes
circular driving
while the second disk species is not passive but experiences a
$y$-direction ac drive,
we show that different
types
of phases can occur depending on whether the ac drive and circular drive
are in or out of phase.
These states include various types of laned structures as well as
fully phase separated states and reentrant reversible phases. 

This work was supported by the US Department of Energy through
the Los Alamos National Laboratory.  Los Alamos National Laboratory
is operated by Triad National Security, LLC, for the
National Nuclear Security Administration of the U. S. Department
of Energy (Contract No. 892333218NCA000001).

\end{document}